\documentclass[preprint]{aastex}
\usepackage{amsmath}
\usepackage{multirow}
\usepackage{geometry}
\usepackage{lscape}
\usepackage{hyperref}
\usepackage{breakurl}
\usepackage{textcomp}
\usepackage{cleveref}[2011/03/22]
\usepackage{fancyhdr}

\crefformat{footnote}{#2\footnotemark[#1]#3}

\newcommand{\HI}{\ion{H}{1}}
\newcommand{\et}{et al.}
\newcommand{\kms}{km~s$^{-1}$}
\newcommand{\s}{$\sim$}
\newcommand{\n}{$-$}

\begin{document}
\slugcomment{\today}

\title{The \HI\ Chronicles of LITTLE THINGS BCDs II: The Origin of IC 10's \HI\ Structure}
\author{Trisha Ashley}
\affil{Department of Physics, Florida International University}
\affil{Current Location: NASA Ames Research Center, Moffett Field, CA 94035}
\email{trisha.l.ashley@nasa.gov}

\author{Bruce G. Elmegreen}
\affil{IBM T. J. Watson Research Center, PO Box 218, }
\affil{Yorktown Heights, New York 10598}
\email{bge@us.ibm.com}

\author{Megan Johnson}
\affil{CSIRO Astronomy \& Space Science}
\affil{P.O. Box 76, Epping, NSW 1710 Australia}
\email{megan.johnson@csiro.au}

\author{David L. Nidever}
\affil{Department of Astronomy, University of Michigan} 
\affil{Ann Arbor, MI, 48109, USA} 
\email{dnidever@umich.edu}

\author{Caroline E. Simpson}
\affil{Department of Physics, Florida International University}
\affil{11200 SW 8th Street, CP 204, Miami, FL 33199}
\email{simpsonc@fiu.edu}

\author{Nau Raj Pokhrel}
\affil{Department of Physics, Florida International University}
\affil{11200 SW 8th Street, CP 204, Miami, FL 33199}
\email{npokh001@fiu.edu}

 \begin{abstract}
In this paper we analyze Very Large Array (VLA) telescope and Green Bank Telescope (GBT) atomic hydrogen (\HI) data for the LITTLE THINGS\footnote{Local Irregulars That Trace Luminosity Extremes, The \HI\ Nearby Galaxy Survey; \url{https://science.nrao.edu/science/surveys/littlethings}} blue compact dwarf galaxy IC 10.  The VLA data allow us to study the detailed \HI\ kinematics and morphology of IC 10 at high resolution while the GBT data allow us to search the surrounding area at high sensitivity for tenuous \HI.   IC 10's \HI\ appears highly disturbed in both the VLA and GBT \HI\ maps with a kinematically distinct northern \HI\ extension, a kinematically distinct southern plume, and several spurs in the VLA data that do not follow the general kinematics of the main disk.  We discuss three possible origins of its \HI\ structure and kinematics in detail: a current interaction with a nearby companion, an advanced merger, and accretion of intergalactic medium.  We find that IC 10 is most likely an advanced merger or a galaxy undergoing accretion.  

\end{abstract}
 
 
\section{Introduction}

Dwarf galaxies are generally inefficient star formers \citep{leroy08}, however, blue compact dwarf (BCD) galaxies are undergoing a burst of star formation.  It is unclear what has triggered the high star formation rates seen in BCDs but not in other dwarf irregular galaxies.  Numerous surveys of BCDs have suggested that the starbursts in BCDs are the results of either mergers with other dwarfs, consumption of intergalactic medium (IGM), or interactions with other nearby galaxies \citep{taylor97, wilcots98, noeske01, pustilnik01, bekki08, koulouridis13, verbeke14}.  Yet there are BCDs that are still thought to be relatively isolated with respect to other galaxies, making mergers and interactions unlikely starburst triggers for these galaxies \citep{taylor97, zitrin09, nicholls11, simpson11, ashley13}.  \citet{Brosch04} suggest that even if a potential companion is present in the vicinity of a BCD, the starburst may not have been triggered by the companion.  Without knowing what has triggered the burst of star formation in BCDs, it is difficult to understand their evolution and model how they are related to other dwarf galaxies.  

In the bottom-up theory of galaxy formation, massive galaxies are thought to have formed from dwarf-like galaxies via mergers and accretion of the IGM.  If BCDs are the result of a merger or accretion of the IGM, then they may be useful analogs to galaxy formation in the early universe.  Signatures of mergers and interactions include tidal tails and bridges \citep{toomre72}.

IC 10 (=UGC 192) is a nearby BCD at a distance of only 0.7 Mpc.  Some basic information on IC 10 can be found in Table~\ref{tab:galinfo}.  IC 10 is the only known starburst galaxy in the Local Group.   It is a distant M31 satellite \citep{ibata13, tully13} and is close enough to have proper motion measurements from masers \citep{brunthaler07}.  Its proximity makes it an excellent subject to study; however, its observational proximity to the Galactic plane makes it difficult to determine extinction.  This difficulty has led to some dispute over IC 10's classification since the colors of the galaxy are dependent upon the extinction of the galaxy.  \citet{richer01} completed a study of IC 10's extinction and concluded that IC 10 should be classified as a BCD.   \citet{hunter04} indicate that its closest companion within $\pm$150 \kms\ is M31 at a relative distance of 250 kpc and velocity difference of 48 \kms.  

The \HI\ of IC 10 has been well studied \citep{huchtmeier79, shostak89, wilcots98, manthey08, nidever13}, revealing an extended \HI\ pool in lower resolution, higher sensitivity data.  The higher resolution, lower sensitivity data reveals a southern plume, three spurs around the galaxy, and many holes/shells.  The \HI\ data also reveal disturbed velocity fields in IC 10: the outer regions of the \HI\ are counter-rotating with respect to the inner gas \citep{shostak89, wilcots98}.  \citet{shostak89} and \citet{wilcots98} suggested that the counter-rotating gas and apparent isolation of IC 10 point to it being a galaxy that is an advanced merger or a galaxy that is still in formation, collecting the primordial gas surrounding it.  \citet{yin10} modeled the chemical evolution of IC 10.  They concluded that slow gas accretion on the timescale of 8 Gyr is consistent with their models. \citet{nidever13} presented a newly discovered \HI\ northern extension in their Robert C. Byrd Green Bank Telescope\cref{note1} (GBT) data.   \citet{nidever13} found that the northern extension is not likely due to stellar feedback, ram pressure stripping, or an interaction with M31, instead, they suggested that the northern extension is the result of an interaction or ongoing merger with an unknown companion.  

IC 10 also has a well-studied stellar component.  \citet{massey95} and \citet{jarrett03} measured a stellar disk diameter of 6\arcmin-7\arcmin, but \citet{sanna10} show that the disk diameter may be 20\arcmin-46\arcmin, possibly larger.  \citet{massey95} also detect an unusually high number of Wolf-Rayet (WR) stars in IC 10; its disk contains at least 15 WR stars while 1 WR star is typical for a star-forming dwarf galaxy.  This indicates that IC 10 is forming massive stars at the same rate as a star-forming region in a spiral galaxy, such as M33, throughout its disk.  \citet{gonclaves12} used the Subaru telescope to study the kinematics of 35 planetary nebulae in IC 10 that traced the stellar motions.  They concluded that the planetary nebulae generally follow the same kinematic trends as the gas, as measured by Very Large Array\footnote{\label{note1}The National Radio Astronomy Observatory is a facility of the National Science Foundation operated under cooperative agreement by Associated Universities, Inc.} (VLA), including the kinematics of some of the outer, counterrotating gas.

\begin{deluxetable}{ccccccc}
\tablecaption{Properties of IC 10\label{tab:galinfo}}
\tabletypesize{\footnotesize}
\tablewidth{0pt}
\tablehead{
\colhead{RA (2000.0)} &  \colhead{Dec (2000.0)} &  \colhead{Distance\tablenotemark{a}} & \colhead{Systemic} & \colhead{$\rm{R}_{\rm{D}}$\tablenotemark{b}} & \colhead{$\rm{log\ SFR}_{\rm{D}}$\tablenotemark{c}} & \colhead{$\rm{M}_{\rm{V}}$\tablenotemark{a}} \\ \colhead{(hh mm ss.s)} & \colhead{(dd mm ss)} & \colhead{(Mpc)} & \colhead{Velocity (\kms)} & \colhead{(kpc)} & \colhead{($\rm{M}_{\sun}\ \rm{yr}^{-1}\ \rm{kpc}^{-2}$)} &  \colhead{(mag)}}

\startdata

00 20 21.9 & 59 17 39 & 0.7 & -348 & $0.40\pm0.01$ & $-1.11\pm0.01$ & $-16.3$ \\  

\enddata

\tablenotetext{a}{\citet{hunter12}}
\tablenotetext{b}{$\rm{R}_{\rm{D}}$ is the V-band disk scale length \citep{hunter06}.}
\tablenotetext{c}{$\rm{SFR}_{\rm{D}}$ is the star formation rate, measured from H$\alpha$, normalized to an area of $\pi \rm{R}_{\rm{D}}^{2}$ \citep{hunter12}}

\end{deluxetable}

In this paper we present VLA data of IC 10 from LITTLE THINGS\footnote{Local Irregulars That Trace Luminosity Extremes, The \HI\ Nearby Galaxy Survey; \url{https://science.nrao.edu/science/surveys/littlethings}}  and GBT data of IC 10 combined from three different surveys.  In section \ref{obs} we discuss the observations and data reduction techniques.  In section \ref{results} we present the results of the observations.  Then in section \ref{disc} we discuss the results and three possible interpretations.  Finally, in section \ref{concl} we make some concluding remarks.

 \section{Observations and Data Reduction}\label{obs}
 
 \subsection{The Very Large Array Telescope}
  LITTLE THINGS is an \HI\ VLA survey of 41 dwarf galaxies.  Each galaxy has high angular ($\thicksim$6\arcsec) and high velocity resolution ($\le$2.6 \kms) \HI\ data and ancillary stellar data.  For more information about the survey see \citet{hunter12}.  Basic VLA observing information for IC 10 can be seen in Table \ref{tab:vobsinfo}.  

\begin{deluxetable}{cccc}
\tablecaption{VLA Observing Information \label{tab:vobsinfo}}
\tabletypesize{\small}
\tablehead{
\colhead{Configuration} &  \colhead{Date Observed} &  \colhead{Project ID} & \colhead{Time on Source (hours)}}
\startdata

B & 08 Jan 15, 08 Jan 21, 08 Jan 27 & AH927 & 11.2\\
C & 08 Mar 23, 08 Apr 15 & AH927 & 5.85\\
D & 08 Jul 8, 08 Jul 24, 08 Jul 25 & AH927 &  0.1\\

\enddata
\end{deluxetable}

The data were calibrated and imaged using a recipe that is discussed in detail in \citet{hunter12}.  Basic information on IC 10's VLA map can be found in Table~\ref{mapinfo}.  The imaging of the VLA data was not done using standard \textsc{clean} in AIPS, instead we implemented a new Multi-Scale (M-S) cleaning algorithm in the AIPS task \textsc{clean}.  M-S \textsc{clean} differs from standard \textsc{clean} by convolving the data to a set of user specified synthesized beam sizes (0\arcsec, 15\arcsec, 45\arcsec, and 135\arcsec\ were chosen for the LITTLE THINGS data).  M-S  \textsc{clean} then finds the region of highest flux amongst all of the convolved data and uses that region to obtain clean components.  This allows us to recover faint emission (from the data convolved with larger beam sizes) with the advantage of still being able to recover high angular resolution details (from the data convolved with the smaller beam sizes).  \citet{cornwell08} discuss the benefits of M-S \textsc{clean} in detail.

\begin{deluxetable}{lcccc}
\tablecaption{\HI\ Map Information\label{mapinfo}}
\tabletypesize{\footnotesize}
\tablewidth{0pt}
\tablehead{
\colhead{Telescope} &  \colhead{Synthesized} &  \colhead{Linear} & \colhead{Velocity Resolution} & \colhead{RMS over \s10 \kms} \\ \colhead{} & \colhead{Beam Size (\arcsec)} & \colhead{Resolution (pc)} & \colhead{(\kms)} & \colhead{(atoms cm$^{-2}$)}}

\startdata

VLA & 8.44 $\times$ 7.45 & 30 & 2.57 & 4.7$\times$10$^{18}$\\ 
GBT &  522 $\times$ 522 & 2000 & 1.61 & 4.8$\times$10$^{16}$\\

\enddata

\end{deluxetable}

\subsection{The Green Bank Telescope}

 IC 10 was observed by three different projects using the GBT.  These three projects later combined their data to increase the signal-to-noise of the maps.  Information on the GBT maps can be found in Table~\ref{mapinfo}.
 
The first project (P.I. Ashley; Proposal ID: GBT13A-430) mapped the \HI\ environments of the six LITTLE THINGS BCDs to look for extended emission and possible perturbers. In this project a 2\degr$\times$4\degr\ field was mapped around IC 10.  The 4\degr\ side of the map was chosen to follow the north-south direction because it was thought that there may be extended emission from the southern plume seen in the VLA data (see section~\ref{results} for more details on the southern plume).  Existing data of a 2\degr$\times$2\degr\ field around IC 10 (P.I. Johnson; Proposal ID: GBT12B-312) were also combined with these data using \textsc{dbcon} in AIPS.  The total time spent observing IC 10 with the GBT in these two projects (including overhead time) was 50.5 hours.  

The data in both these projects were taken using on-the-fly mapping at the Nyquist rate at a velocity resolution of 0.16 \kms. With a 12.5 MHz bandwidth, in-band frequency switching (central frequency switch of 3.5 MHz) was used for preliminary calibrations.  After observations, stray radiation was removed using a program written by NRAO staff.   Standard calibration was then done using GBTIDL's\footnote{Developed by NRAO; documentation is located at  \url{http://gbtidl.sourceforge.net}.} \textsc{getfs}.  The data also had RFI spikes removed by linear interpolation using the neighboring channels.  The data were boxcar smoothed to a velocity resolution of \s1.6 \kms.   Finally, third-order polynomials were fit to emission-free channels to remove residual baselines.  

The third project (P.I. Nidever) conducted a $\sim$300 degree$^2$ survey of the tip of the Magellanic Stream (MS) to map its emission across the Milky Way midplane (Proposal IDs: GBT10B-035 and GBT11B-082; Nidever et al., in prep.). The survey region included IC~10.  The reduction process is detailed in \citet{nidever13} and summarized here.  On-the-fly mapping was used (scanning in right ascension) to obtain frequency-switched, 21-cm spectral line data with a velocity resolution of 0.32 \kms.  Standard calibration was done using GBTIDL's \textsc{getfs}.  The data were binned to a velocity resolution of $\sim$1.6 \kms\ and then the baseline was removed with a custom routine.  The final spectra in the region around IC~10 were resampled onto a rectangular grid in Galactic coordinates at 4$\arcmin$ spacing.  Finally, all of the datasets from the three projects were resampled onto the same Galactic coordinate grid and combined with exposure time weighting.

\section{Results}\label{results}
\subsection{Stellar Component}

\begin{figure}[!ht]
\centering
\epsscale{1.1}
\plottwo{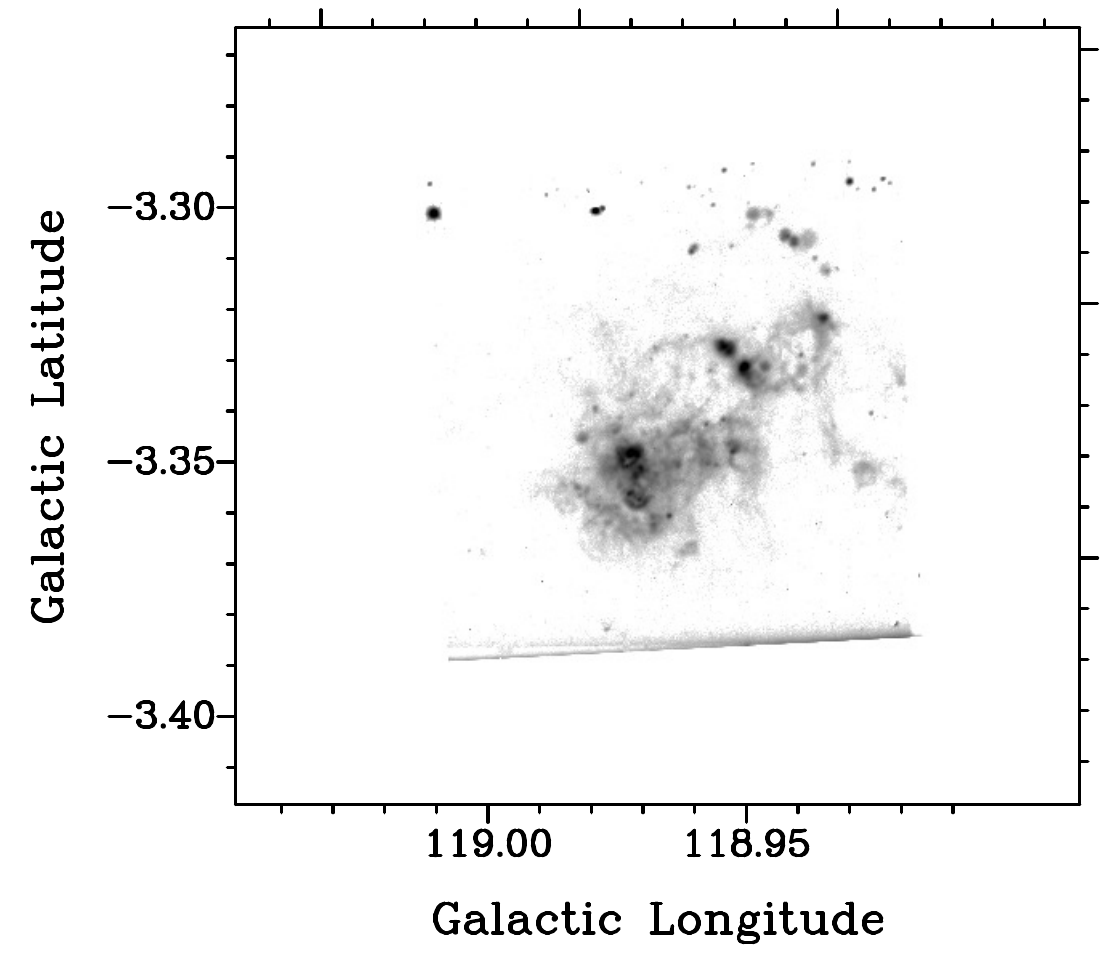}{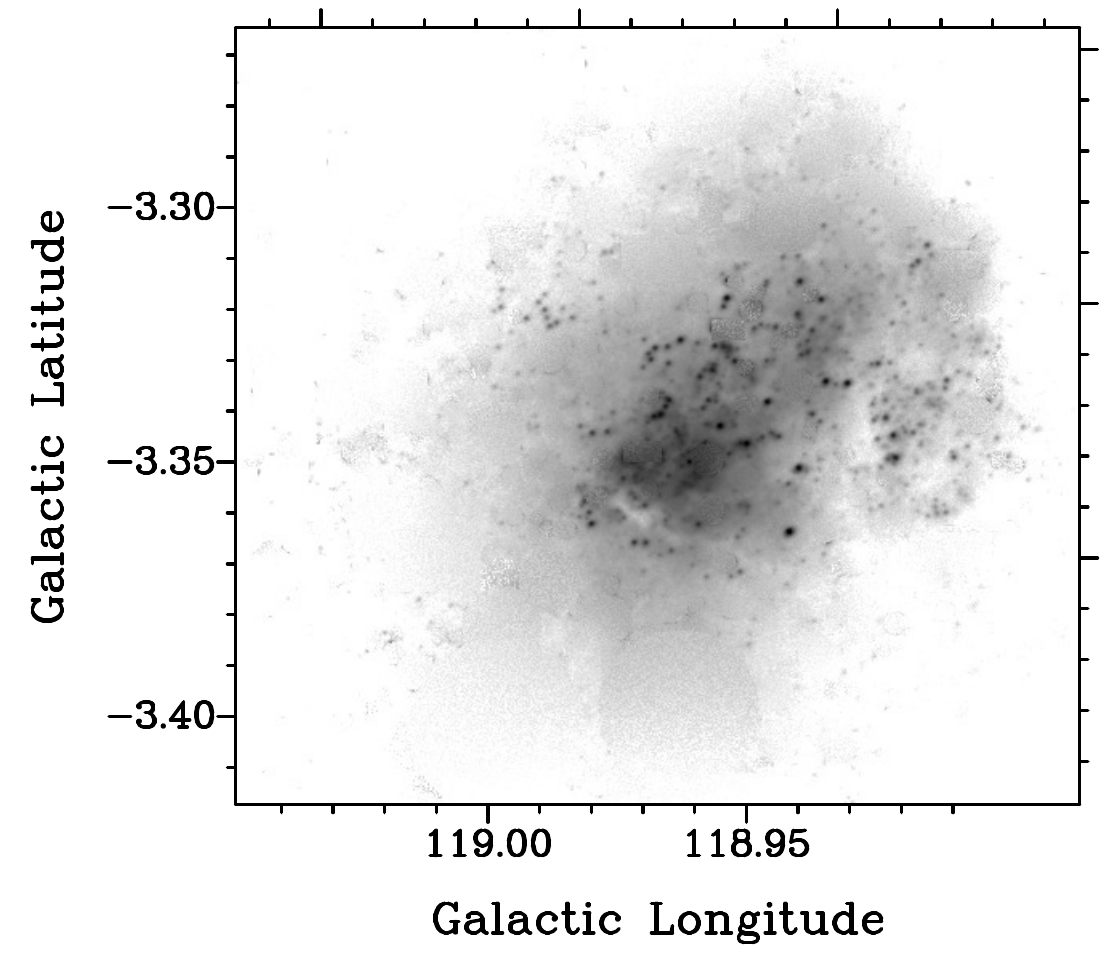}
\caption[IC 10: H$\alpha$ and V-band]{IC 10. \ \  \textit{Left:} H$\alpha$;\ \  \textit{Right:} V-band. \label{ic10_star}}
\end{figure}

IC 10's H$\alpha$ and V-band maps are shown in Figure~\ref{ic10_star}.  The V-band data were taken with the 1.1 m Hall Telescope at Lowell Observatory and represent the star formation that has taken place over the past Gyr \citep{hunter06}.  The H$\alpha$ data were collected with the 1.8 m Perkins Telescope at Lowell Observatory and represents star formation that has occurred over the past 10 Myr \citep{hunter04}.

\subsection{VLA \HI\ Morphology}
\begin{figure}[!Ht]
\epsscale{0.48}
\plotone{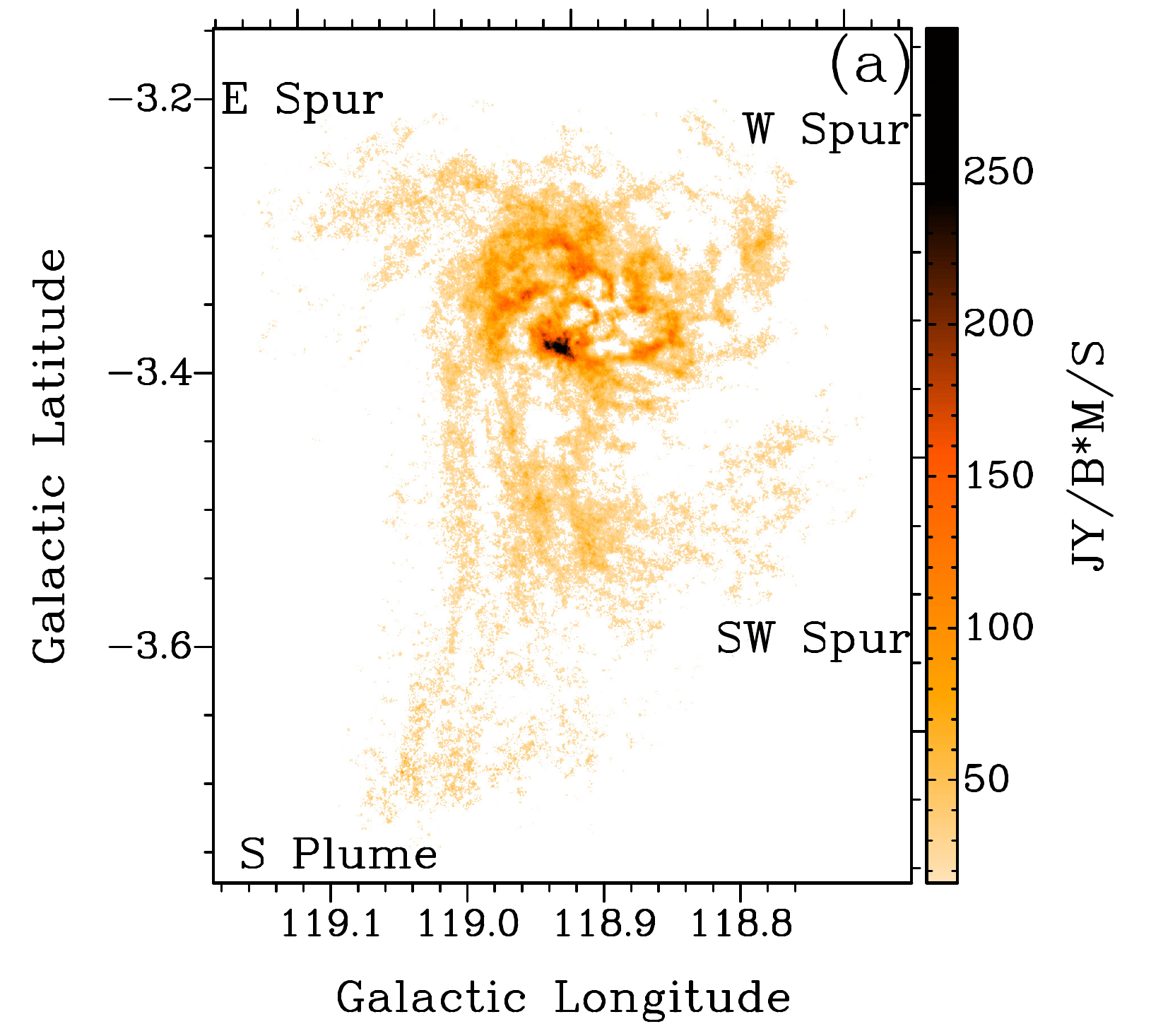}
\epsscale{0.497}
\plotone{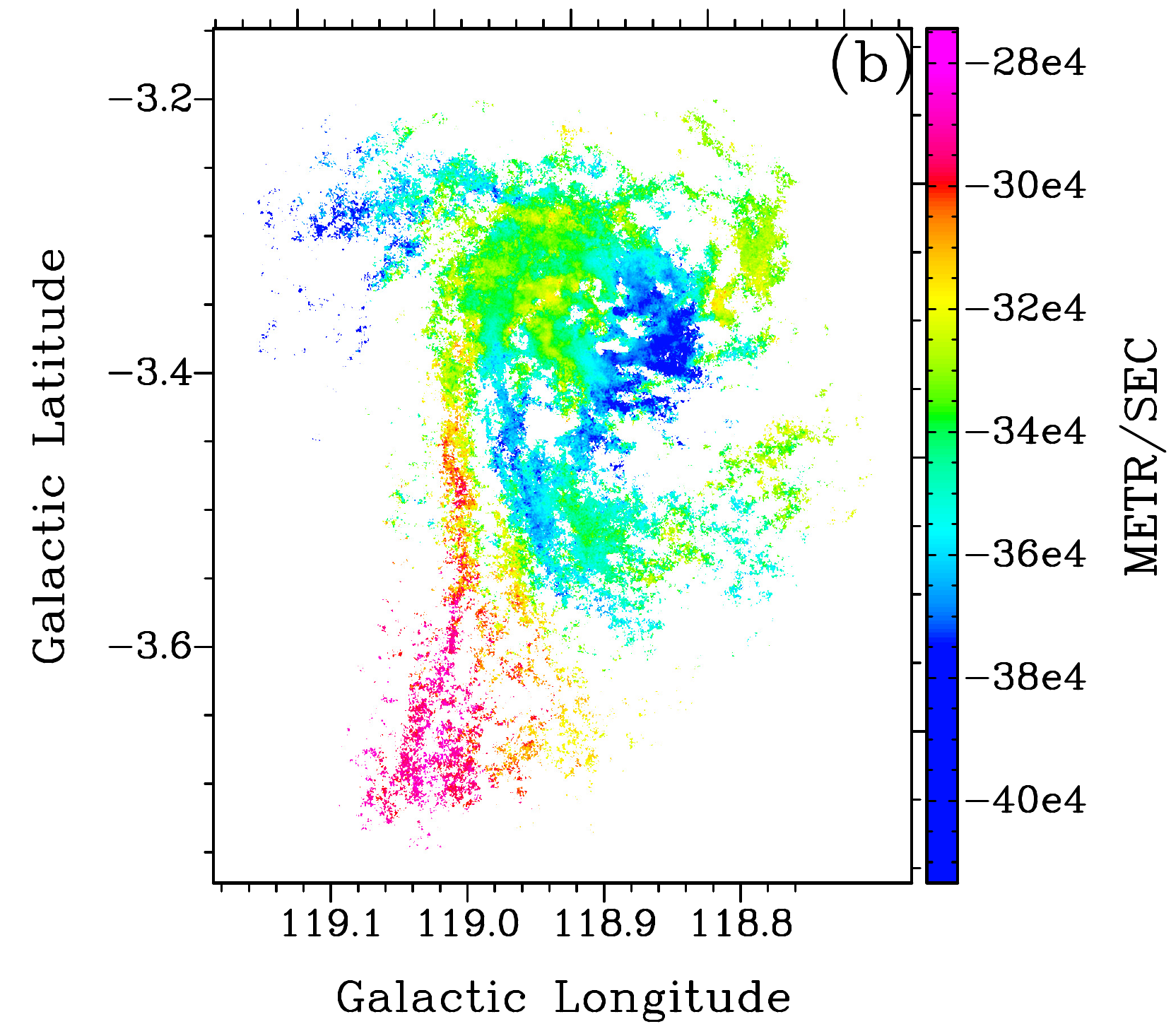}
\epsscale{0.48}
\plotone{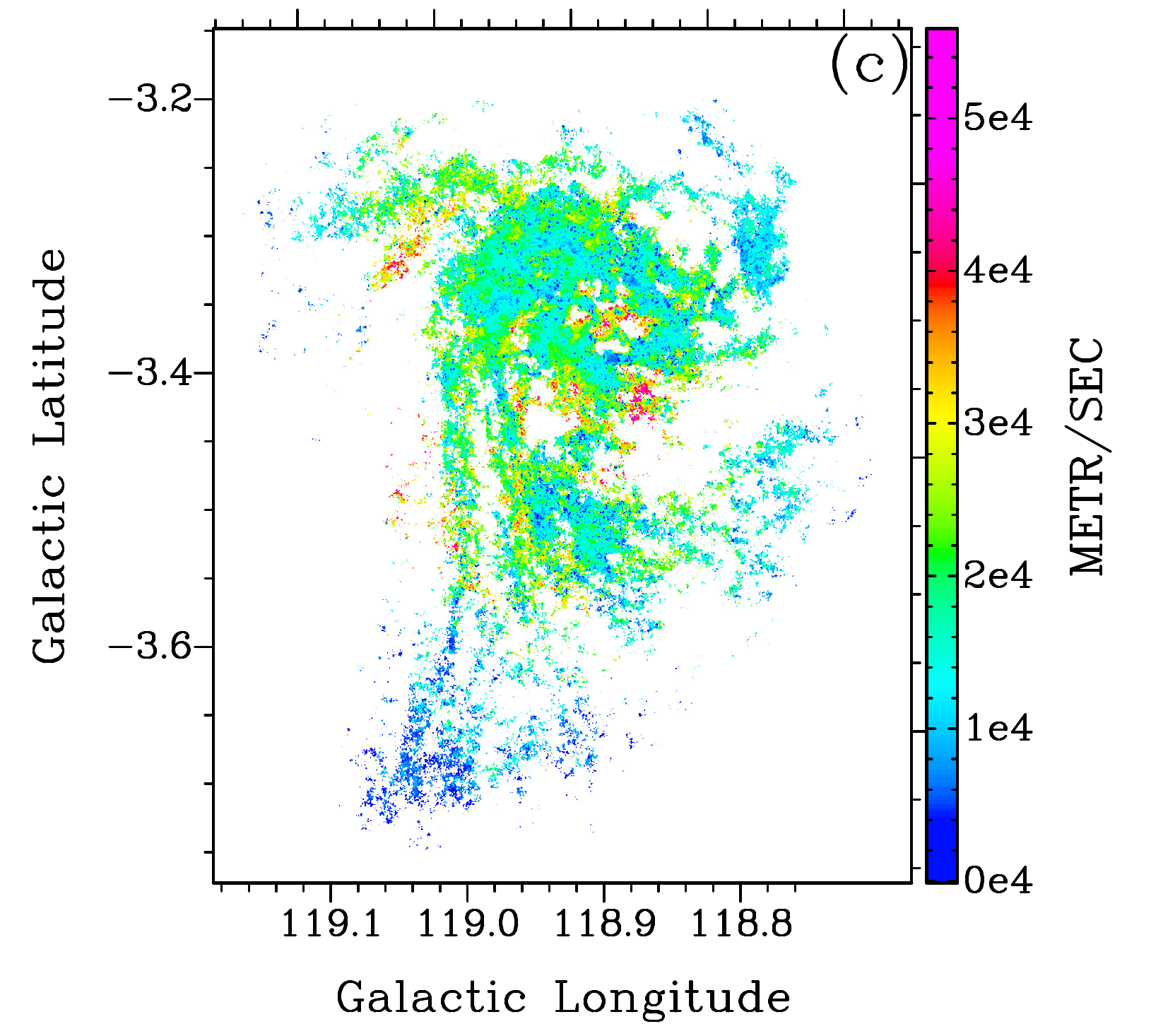}
\caption[IC 10: VLA integrated \HI\ natural-weighted moment maps]{IC 10's natural-weighted moment maps from VLA observations. (a): Integrated \HI\ intensity map blanked at a 2$\sigma$ level where 1$\sigma=14.7\times10^{19}\ \rm{atoms}\ \rm{cm}^{-2}$.  (b): Intensity weighted velocity field. (c): Velocity dispersion field. \label{ic10vla}}
\end{figure}

\begin{figure}[!Ht]
\epsscale{1.12}
\plottwo{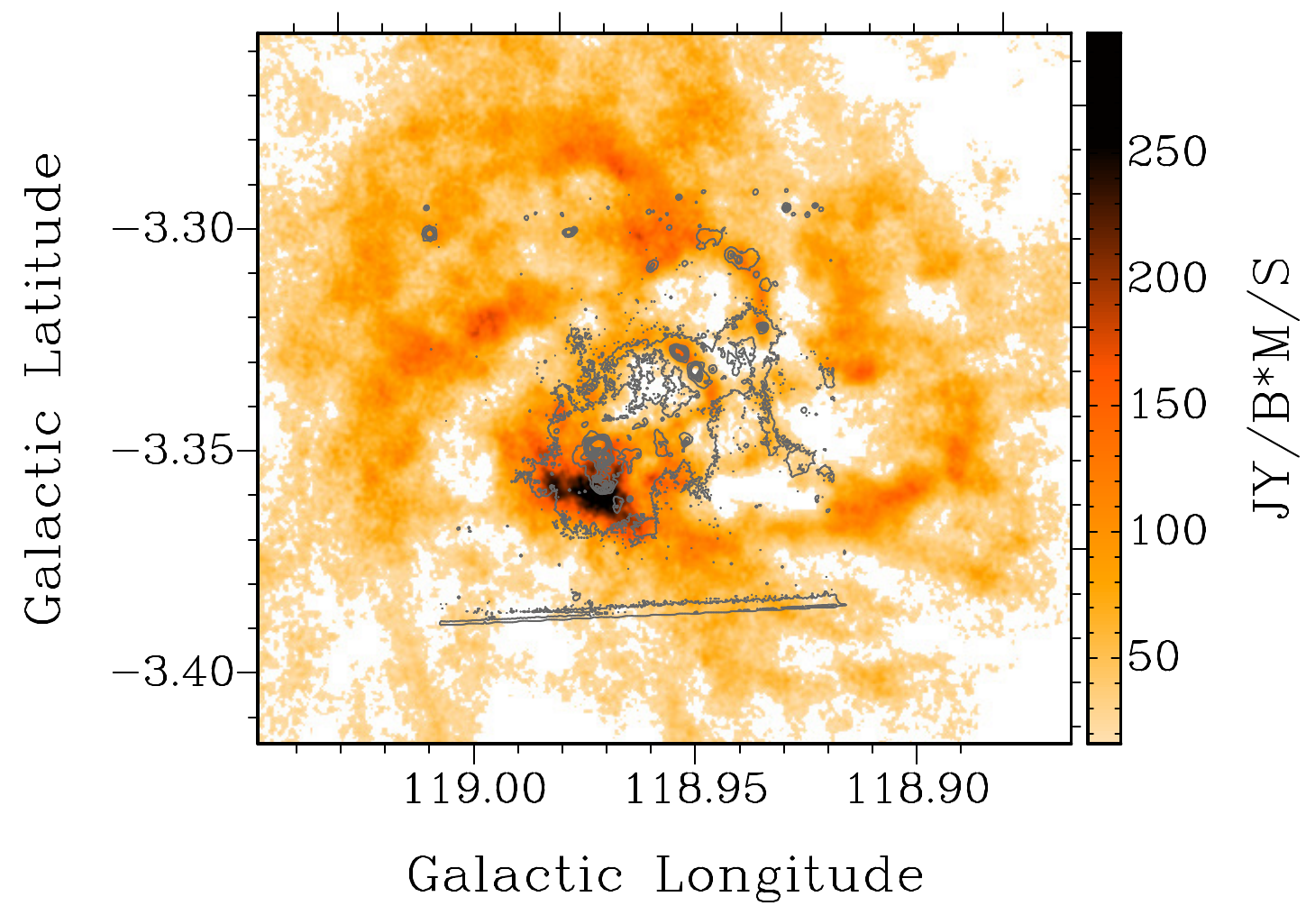}{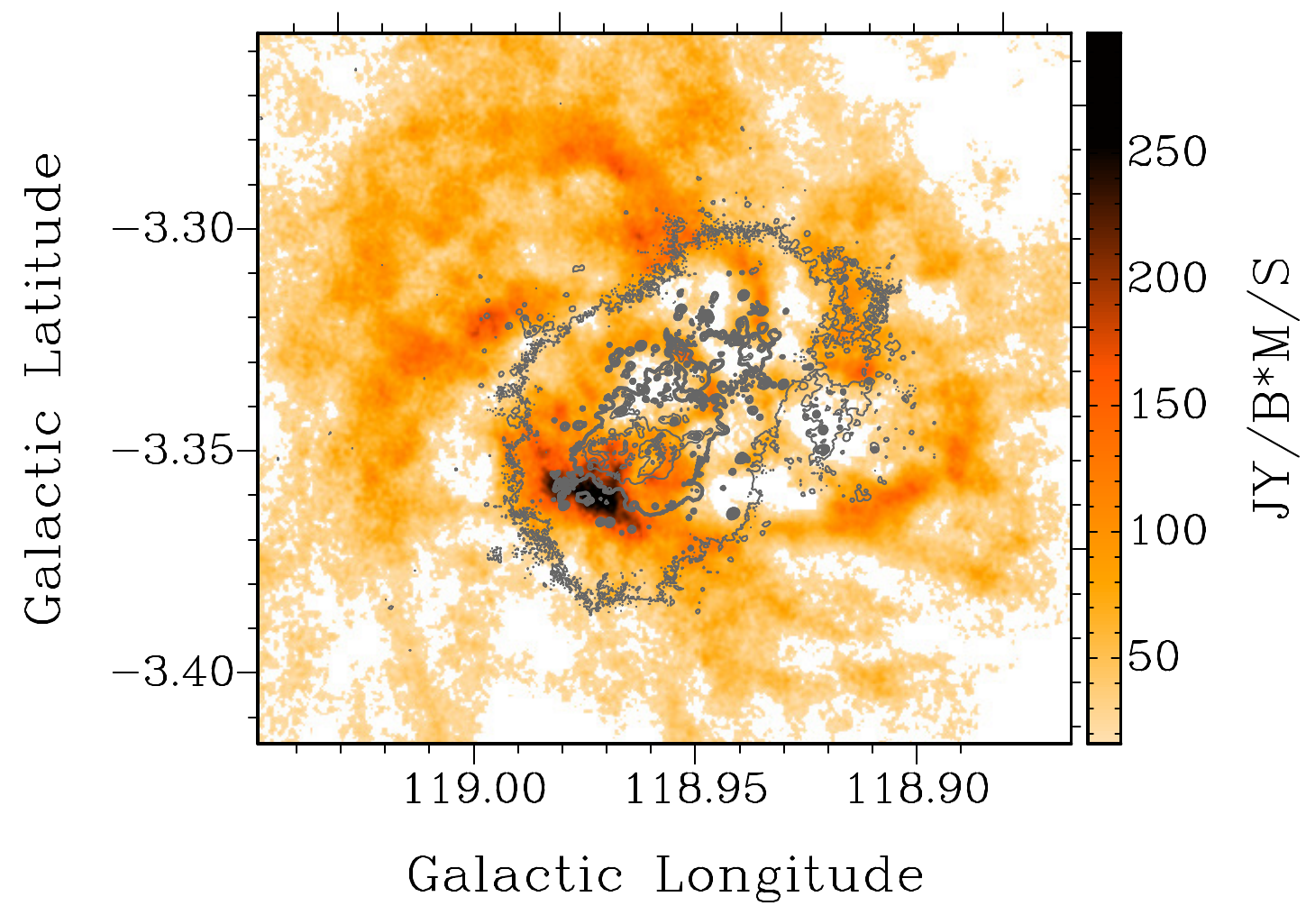}
\caption[IC 10: VLA integrated \HI\ natural-weighted map overlaid with stellar contours]{A colorscale map of IC 10's natural-weighted, main disk \HI\ column density overlaid with stellar contours (note this map is on a different scale than those in Figure~\ref{ic10vla}): \textit{Left:} H$\alpha$. \textit{Right:} V-band. \label{ic10starkntr}}
\end{figure}

The natural weighted, integrated \HI\ intensity map as measured by the VLA is shown in Figure~\ref{ic10vla}a.  \HI\ features that have been pointed out by previous authors \citep{wilcots98, manthey08} are the southern plume, the southwestern spur, the western spur, and the eastern spur.  The main disk of IC 10 is often considered to be the circular region of gas between all of these features.  Figure~\ref{ic10starkntr} shows a zoomed in image of the main body of IC 10 with the colorscale of the natural weighted \HI\ column density and the contours of the stellar components.

\subsection{VLA \HI\ Velocity and Velocity Dispersion Field}
The VLA \HI\ velocity field is shown in Figure~\ref{ic10vla}b.  The main disk of IC 10 (not including the spurs or southern plume) has some regular rotation where the blueshifted velocities are in the southwest and the redshifted velocities are in the northeast.  The spurs do not follow this velocity trend; they have velocity changes in opposite directions to that of the main disk.  For example, the gas in the eastern spur becomes blueshifted toward the northeast.  The southern plume appears kinematically distinct from the rest of the galaxy, with velocities reaching \s70 \kms\ less negative than IC 10's systemic velocity.  The velocity dispersions, shown in Figure~\ref{ic10vla}c, vary greatly and are high ($\ge$20 \kms)  throughout most of the galaxy \citep[10 \kms\ is typical for an undisturbed dwarf galaxy:][]{tamburro09, warren12}.

\subsection{GBT \HI\ Morphology}
\begin{figure}[!Ht]
\centering
\epsscale{0.96}
\plottwo{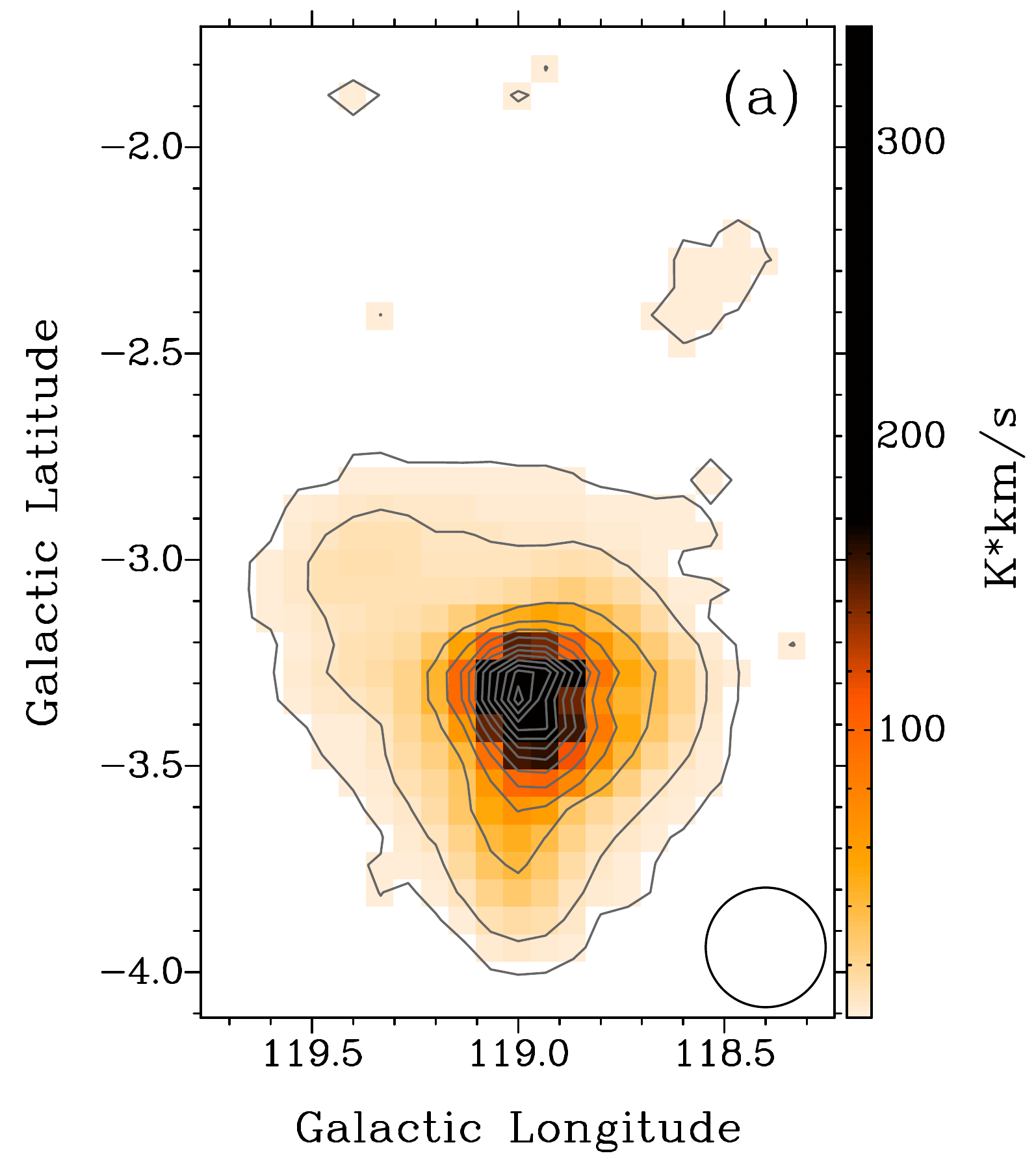}{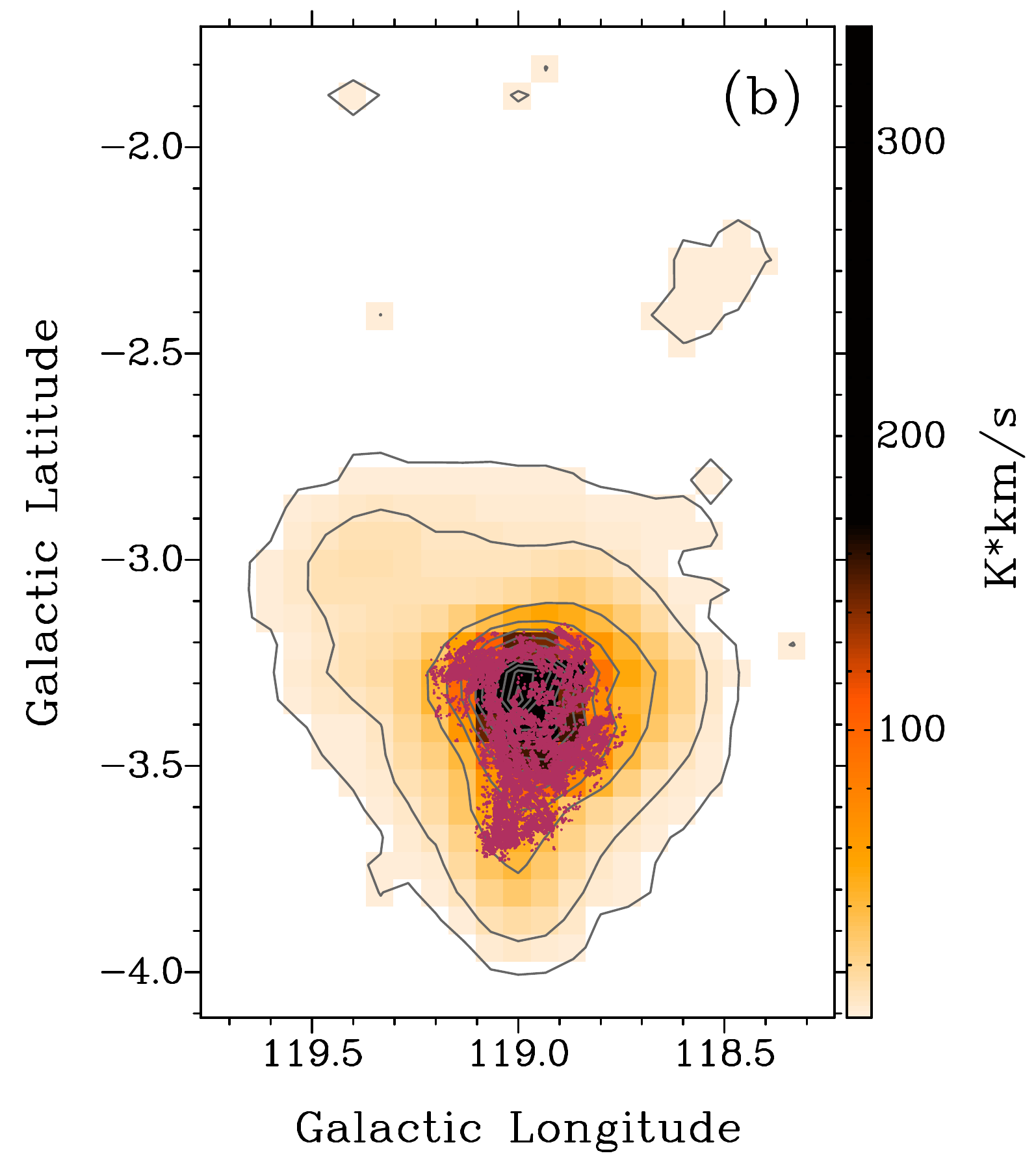}
\epsscale{0.98}
\plottwo{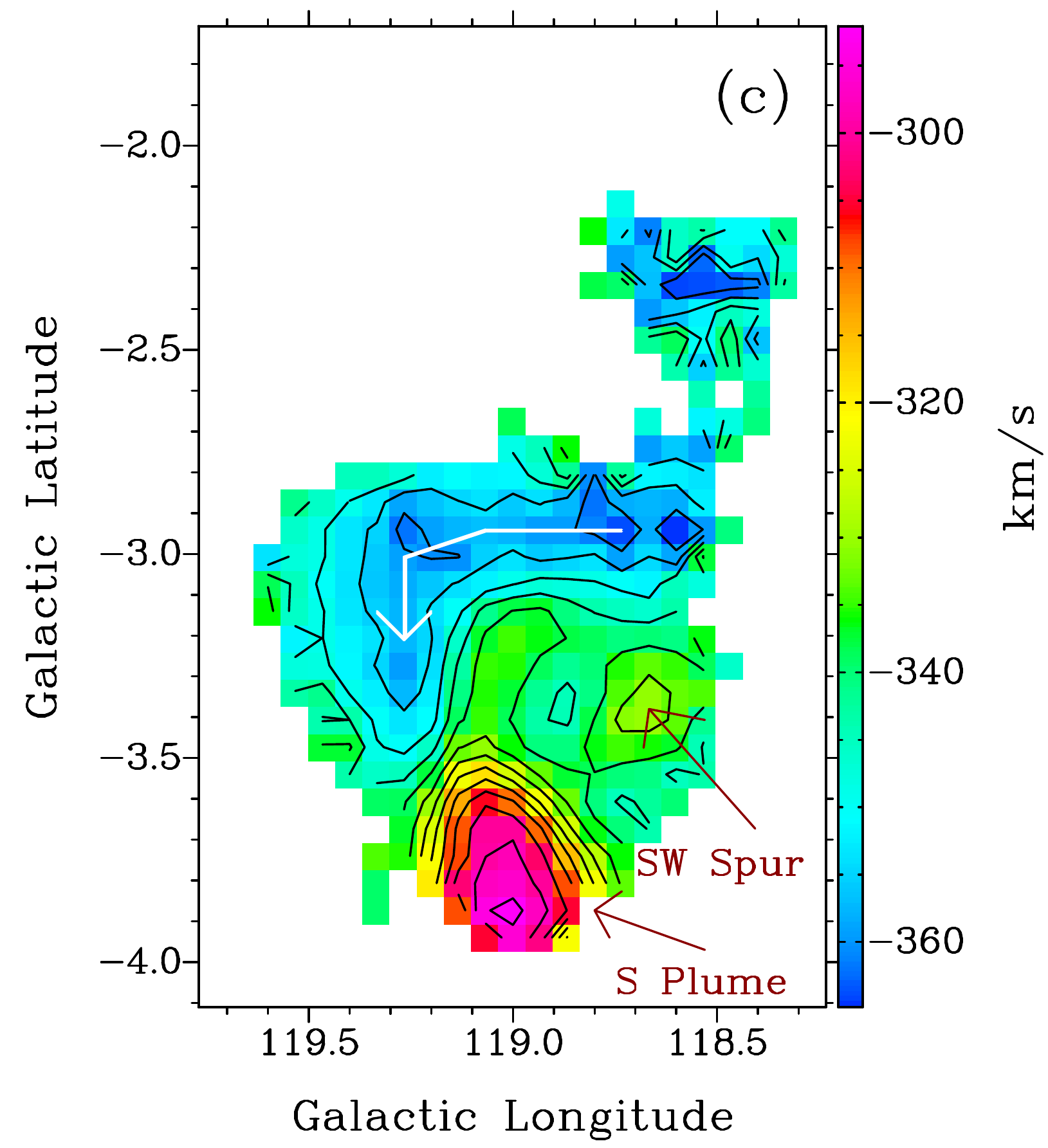}{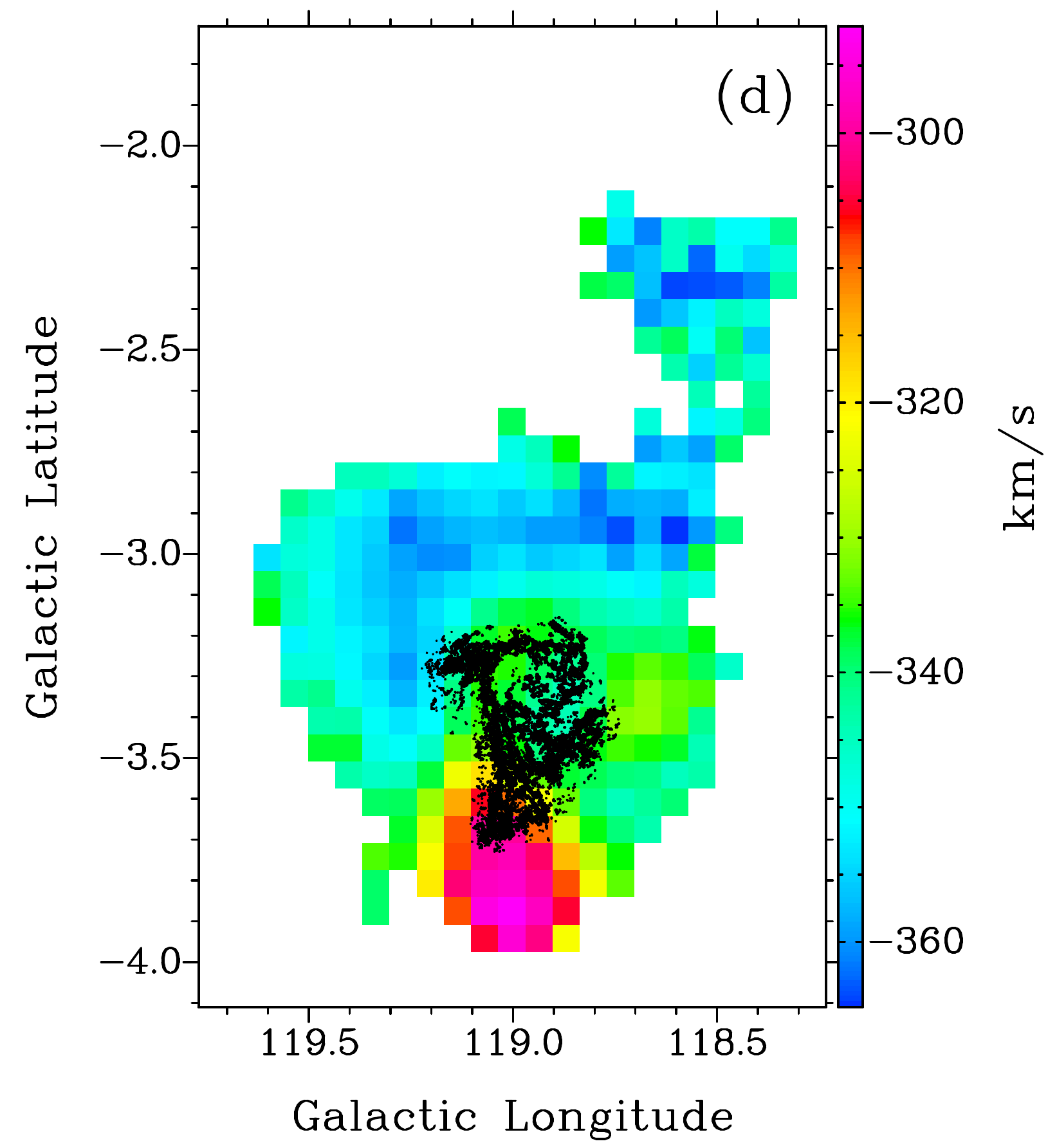}
\caption[IC 10: GBT full integrated \HI\ natural-weighted moment map]{IC 10.\ \  (a): GBT integrated \HI\ intensity map; contour levels are 1$\sigma\times$(9, 30, 130, 230, 330, 430, 530, 630, 730, 830, 930, 1030, 1130) where 1$\sigma=2.5\times10^{17}\ \rm{atoms}\ \rm{cm}^{-2}$. The beam size is shown by the black circle in the bottom right of the map.  \ \  (b): GBT integrated \HI\ intensity map with the 2$\sigma$ contour of the VLA natural weighted column density map in red (dark grey in the printed version).  (c): GBT \HI\ velocity field.  Contours are \n360 \kms\ to \n294 \kms\ separated by 5.5 \kms.  (d): The GBT velocity field with the 2$\sigma$ contour of the VLA natural weighted column density map in black.  \label{ic10gbt}}
\end{figure}

\begin{figure}[!Ht]
\centering
\epsscale{0.47}
\plotone{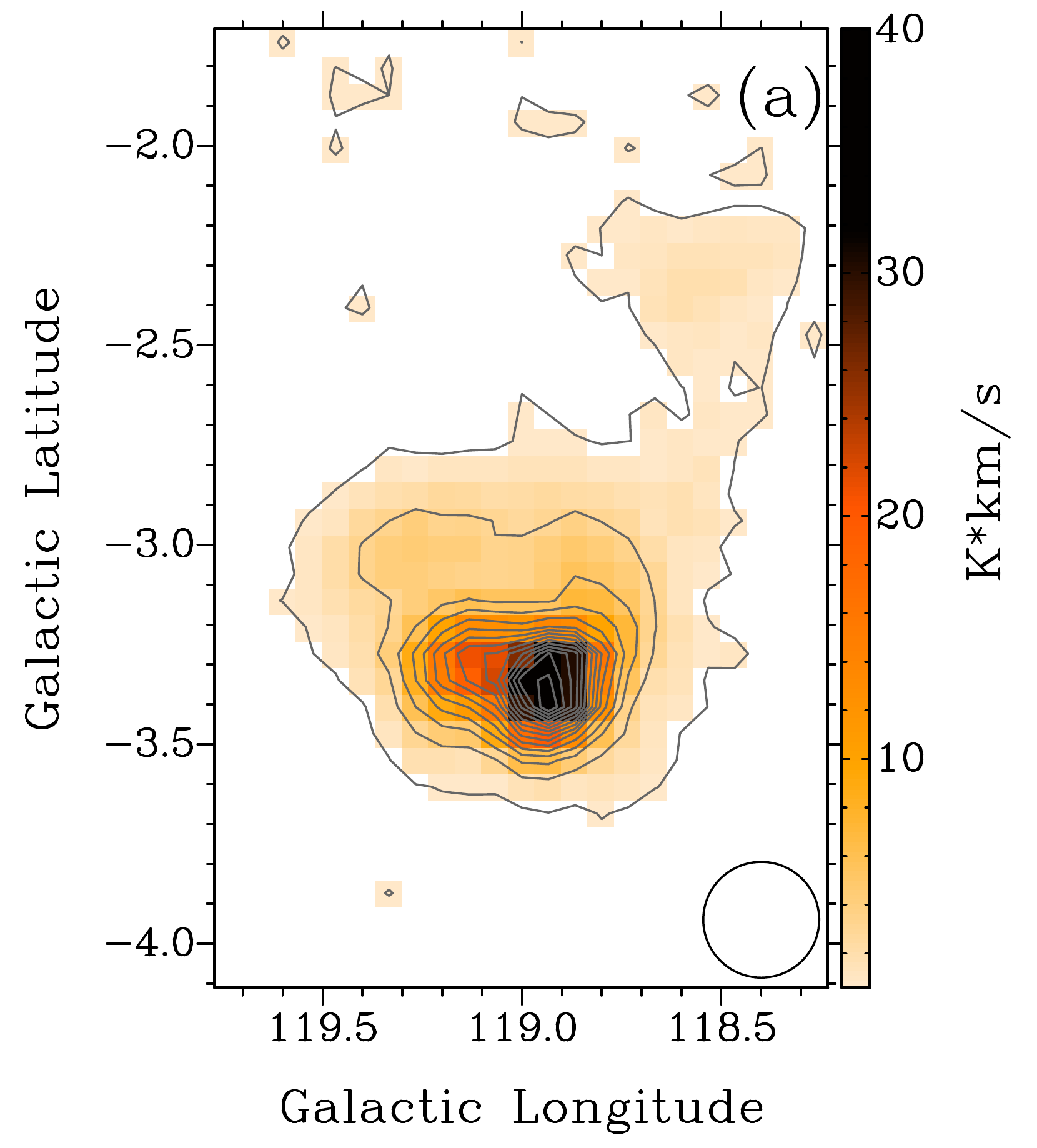}
\epsscale{0.49}
\plotone{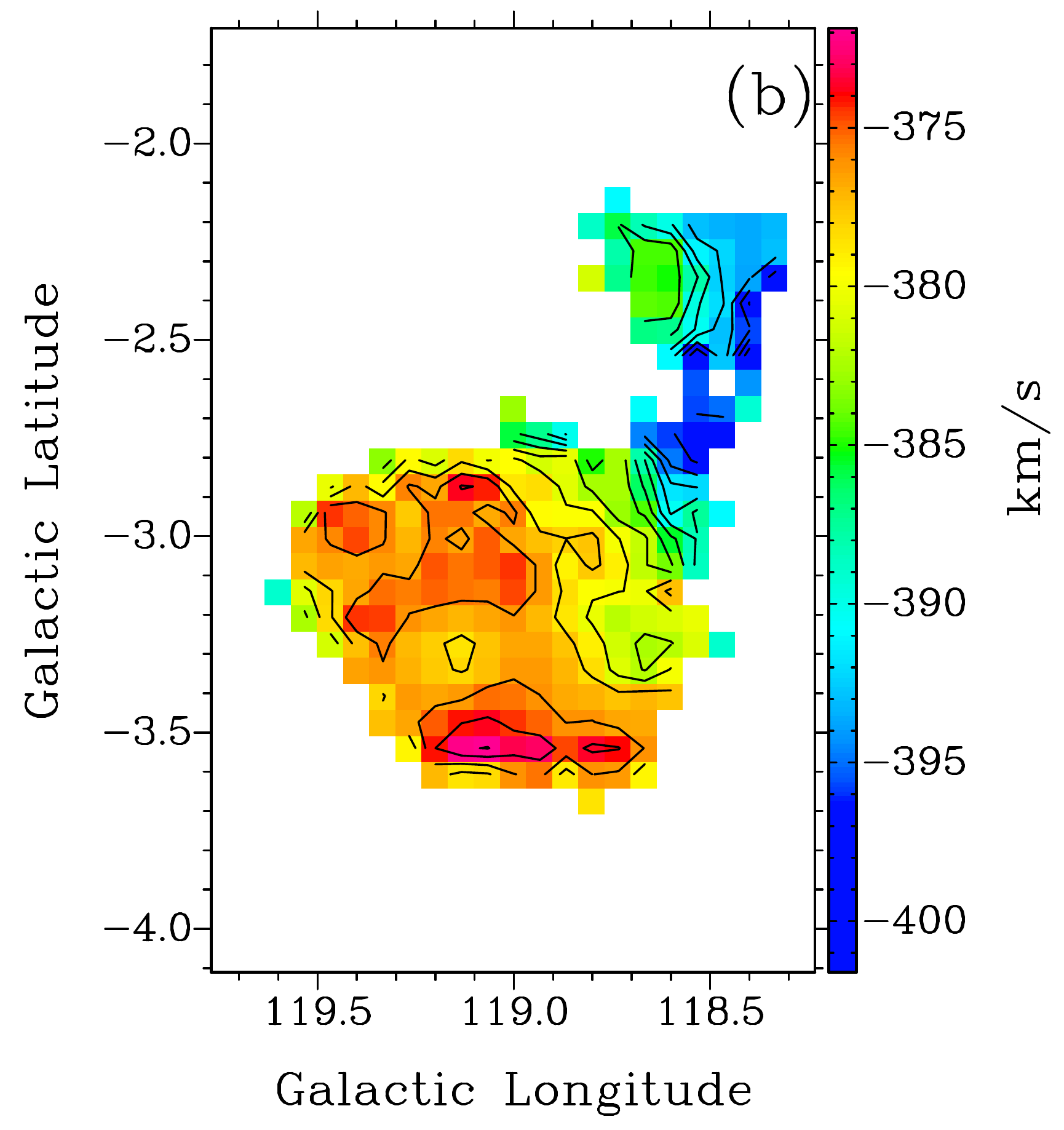}
\caption[IC 10: GBT integrated \HI\ natural-weighted moment map showing the northern extension]{IC 10.\ \  (a): GBT integrated \HI\ intensity map of the northern extension; the velocity range used for integration is \n406.8 to \n373.0 \kms.  Contour levels are 1$\sigma\times$(5, 25, 45, 65, 85, 105, 125, 145, 165, 185, 205, 225, 245, 265, 285) where 1$\sigma=1.2\times10^{17}\ \rm{atoms}\ \rm{cm}^{-2}$. \ \  (b): GBT \HI\ velocity field of the northern extension map.  Contours are \n400 to \n372 \kms\ separated by 2 \kms. \label{ic10gbt_ext}}
\end{figure}

The combined GBT observations of IC 10 cover a 2\degr$\times$4\degr\ area.  The integrated \HI\ intensity map as measured by the GBT is shown in Figure~\ref{ic10gbt}a (this \HI\ column density map of IC 10 was made using the channel maps that cover a velocity range of \n451.16 to \n227.35 \kms).  The size of the beam is shown by the black circle in the bottom right of this map.  This first map is shown again in Figure~\ref{ic10gbt}b with the 2$\sigma$ contour of the VLA data plotted on top for comparison.  Another GBT integrated \HI\ intensity map is shown in Figure~\ref{ic10gbt_ext}, where only the channel maps covering a velocity range of \n406.8 to \n373.88 \kms\ have been integrated.  This velocity range was chosen to highlight a faint extension to the north of IC 10's main body.  This northern extension is the new feature that is presented in \citet{nidever13}.  It has a projected length of 7 kpc, where the endpoints were taken from the most northern tip of the 5$\sigma$ contour to where the disk of the main body of IC 10 and the extension meet in Figure~\ref{ic10gbt_ext} at \s(118.6\degr, \n2.8\degr); the beam is also shown as a black circle in the bottom right of this Figure.

\subsection{GBT \HI\ Velocity Field}

\begin{figure}[!Ht]
\centering
\epsscale{0.7}
\plotone{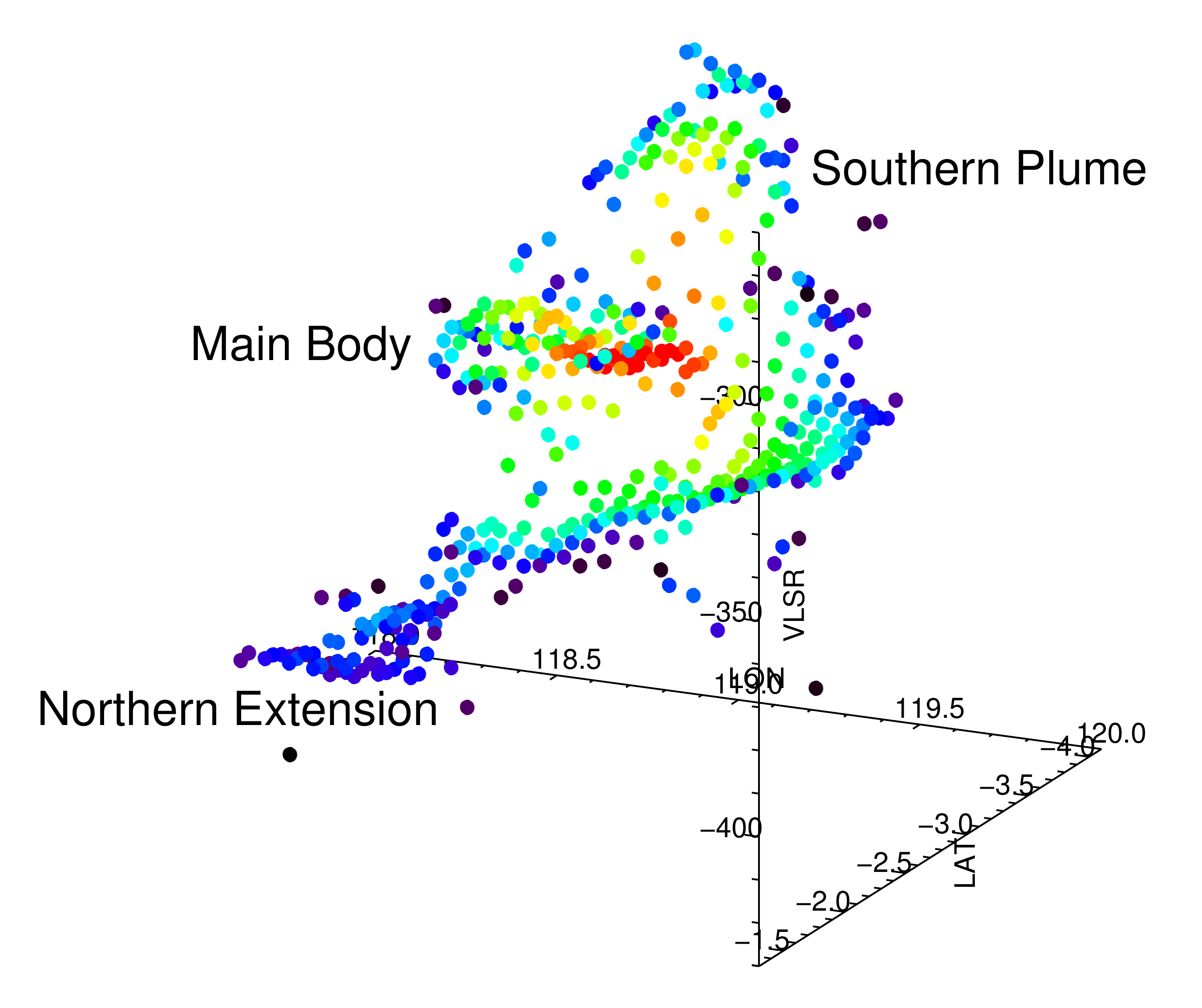}
\caption{A single frame from the animation showing the distribution of \HI\ Gaussian components in Galactic longitude, latitude and $V_{\rm LSR}$ (vertical axis) color-coded by log N(\HI ) (red is high column density).  The full animation illustrates clearly the connection of the northern extension to the counter-rotating gas surrounding IC 10 and can be watched in full in the online version.  \label{ic10_davidsvideo}}
\end{figure}

The full velocity field map shown in Figure~\ref{ic10gbt}c was made by creating a mask from the emission in both Figure~\ref{ic10gbt}a and Figure~\ref{ic10gbt_ext}a, and then integrating using the same channel range as for the column density map shown in Figure~\ref{ic10gbt}a.  A second map is shown in Figure~\ref{ic10gbt}d of the colorscale of the full velocity field with the 2$\sigma$ contour of the VLA \HI\ column density for comparison.  The southern plume (the yellow-pink region in the velocity field centered near 119.0\degr, \n3.9\degr: see the online version for this figure in color) has a distinct velocity range from the rest of the galaxy in the GBT velocity field as it did in the VLA velocity field.  Two regions, one near (118.7\degr, \n3.4\degr) and one near (119.0\degr, \n3.2\degr), show velocity peaks at about \n330 and \n340 \kms\ respectively.  These two features appear to be associated with the east side of IC 10's main disk and southwestern spur mapped by the VLA.  The northern extension also has a distinct velocity range.   Its velocities transition into the main body in the full velocity field in Figure~\ref{ic10gbt}c, starting in the northwest of the main body, then over to the northeast, and finally down the east side of the main body (traced by the white arrow in the northeast of the velocity field in Figure~\ref{ic10gbt}c).  The northern extension occurs at velocities almost 200 \kms\ away from the Milky Way emission that appears in the GBT data cubes.  This indicates that it is not likely a foreground gas cloud in the Milky Way. It is possible that the northern extension is a high velocity cloud related to the Milky Way, however, the smooth transition of velocities into the main body of IC 10 indicate that this northern extension is likely associated with IC 10.   This transition of velocities can be seen more clearly in Figure~\ref{ic10_davidsvideo} (Figure~\ref{ic10_davidsvideo} is a selected frame from a video that can viewed fully online).  Figure~\ref{ic10_davidsvideo} shows the result of a Gaussian decomposition of our IC 10 GBT datacube using the software described in \citet{nidever08}; IC 10's main \HI\ disk is seen in red at the center of the image, while the southern plume extends upwards (more positive velocity) and the northern extension downwards (more negative velocity) in the image.   In this image the northern extension's velocities can be seen transitioning smoothly along the velocity axis to the velocities of IC 10's main body.   The selected velocity range that was used to highlight the northern extension was used to make a velocity field map shown in Figure~\ref{ic10gbt_ext}b.  The part of the northern extension that is furthest north has a round appearance (possibly due to the beam) with a small velocity change of \s10 \kms\ in the east-west direction across the body.  The emission below the northern extension in Figure~\ref{ic10gbt_ext} is from both the northern extension as it transitions into IC 10's main body and IC 10's main disk in the VLA maps (it is also likely that the eastern spur emission is contributing to this part of the velocity field) and these separate features cannot be disentangled from this map alone.

\subsection{\HI\ Mass}
The total \HI\ mass from the VLA data is measured to be $5\times10^{7}M_{\sun}$ and the total mass of the GBT data was measured to be $8\times10^{7}M_{\sun}$.  The VLA was able to recover 60\% of the GBT flux; this difference is reasonable considering the large amount of large scale structure and tenuous emission surrounding IC 10.  Interferometers (like the VLA) act as spatial filters, and only structures smaller than a certain angular size can be detected.  The largest angular size detectable by the VLA (at 1.5 GHz with the most compact configuration) is approximately 16\arcmin.  The VLA is also not as sensitive to faint emission as the GBT.  The combination of these means that the large scale diffuse emission surrounding IC 10 are not detected in the VLA data.  The mass of the northern extension is also measured to be $6\times10^{5}M_{\sun}$ or 0.8\% of the total GBT mass.  The area used to measure the mass of the northern extension is where the northern extension meets the main body of IC 10 at \s(118.6\degr, \n2.8\degr) in the GBT integrated \HI\ intensity map to the top of the 5$\sigma$ contour level of the \HI\ map in Figure~\ref{ic10gbt_ext}.

\section{Discussion}\label{disc}
The morphology and kinematics of the gas in IC 10 is complex as shown in both the GBT and VLA images.  The new extension seen in the GBT \HI\ data may reveal what has happened to IC 10 in the past.  In this section we will discuss three possible explanations for IC 10's \HI\ morphology and kinematics in detail.  These three explanations were chosen for discussion since they are the clearest, simplest possible explanations for the \HI\ morphologies and kinematics in IC 10 (a fourth possible explanation is discussed briefly in Section~\ref{concl}).

\subsection{A Galaxy with a Northern Bridge}\label{bridge_explanation}
\begin{figure}[!Ht]
\centering
\epsscale{0.45}
\plotone{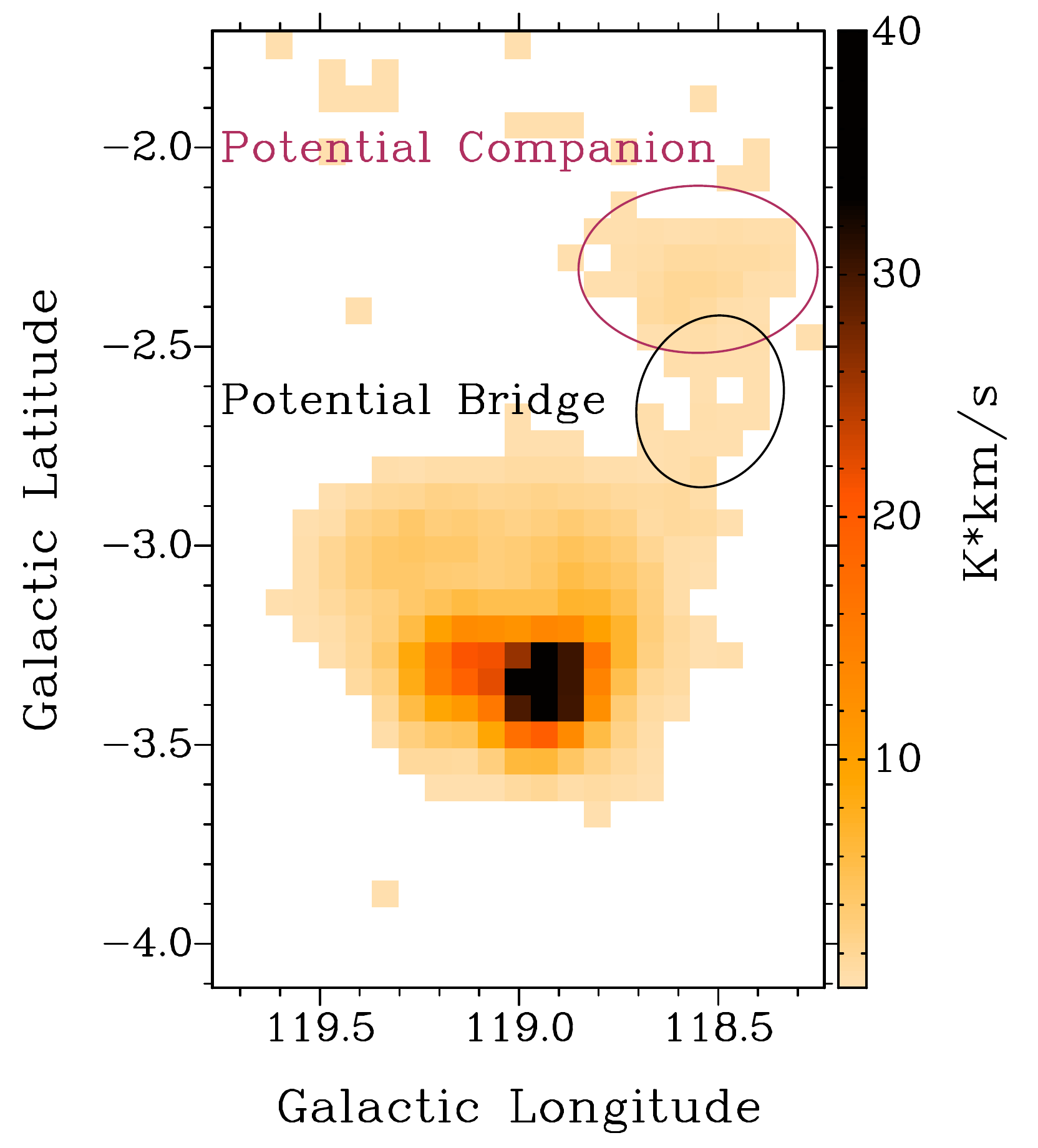}
\caption{A labeled version of Figure~\ref{ic10gbt_ext} showing the location of the features in the case that the northern extension is a bridge and companion galaxy. \label{ic10_bridgelabel}}
\end{figure}

One possible explanation for the northern extension is a bridge, where the round region at the end of the northern extension is a companion to IC 10 (see Figure~\ref{ic10_bridgelabel}).  If the round region at the end of the northern extension is a companion, then it would likely have a stellar component.  DSS\footnote{The Digitized Sky Survey http://archive.stsci.edu/cgi-bin/dss\_form} and NED\footnote{NASA/IPAC Extragalactic Database (NED) http://ned.ipac.caltech.edu/} show no known stellar component in the end of the northern extension.  This does not exclude the round region as a companion; the stellar component may be too faint to be picked up in DSS.

If part of the northern extension is a bridge, then it could be material that has been pulled from the outer disk of IC 10 or the outer disk of the potential companion.  First we will explore the possibility that the northern extension is partly a bridge that is made from material pulled from IC 10's main disk. We first note that in Figure 4c, the potential bridge material appears to transition into IC 10's main body smoothly in projected velocity along an arc that wraps \textit{outward} away from IC 10's main disk in a clockwise direction (see the arrow in Figure~\ref{ic10gbt}c and the green dots in Figure 6 on the online version that lie between the main disk and the northern extension).  Most of the arc is north of the HI emission seen in the VLA map of Figure 2. If this arc were made by the same interaction that made the northern extension, then that implies that as the companion moved away, it pulled material from the main disk, leaving an arc of gas that increases in radius from east to west trailing behind it. We then wonder if such motions are likely given the structure of the eastern spur.  The eastern spur in the VLA map, which extends to about (119.2\degr, \n3.3\degr) and has a velocity range of 380 to 360 \kms, overlaps with this arc. The arc at the same location has velocities of \s360 \kms, indicating that the eastern spur could be part of the northern extension and therefore part of the bridge.  However, if the northern extension and the eastern spur are both material that originated from the main disk of IC 10, then the orientation of the eastern spur seems wrong for this. Whereas the arc in the north spirals clockwise with increasing radius, the eastern spur spirals counter clockwise with increasing radius, like the other spurs of IC 10.  We conclude that the eastern spur and the arc are likely the same \HI\ feature because of their overlapping location in the line of sight and overlapping velocities in the line of sight, however, if they were both part of a bridge that is between IC 10's main disk and a companion to the north, then they could not both have originated from IC 10's main disk.  Therefore, if there is a bridge between IC 10 and a companion to the north, the bridge material did not likely originate from IC 10's main disk.  

Additionally, the velocity fields of the VLA and GBT data do not support the theory that the potential bridge originated from the main disk of IC 10.  If the northern extension is a bridge beginning from IC 10's main disk, that means that it is material that has been pulled off of IC 10's main disk.  In this case, the bridge should rotate like the main disk.  The spurs, as mentioned above, could be part of the bridge material wrapping  around the main disk.  The spurs, however, have velocities that are opposite to the main disk: the eastern spur is blueshifted on the redshifted side of the main disk while the western and southwestern spurs are redshifted relative to the systemic velocity of IC 10 on the blueshifted side of the main disk.  The southwestern spur also has its own velocity gradient opposite of the main disk in the line of sight (redshifted to the west and closer to the systemic velocity of IC 10 in the east).  The rotation of the spurs having the opposite sense of rotation to the disk could be due to IC 10 having a face-on disk, while the spurs could be warps in the disk leading them to appear as if they are counter-rotating to the disk in the line of sight.  However, this seems unlikely because IC 10's main disk in the VLA maps has a change in velocity of \s50 \kms, indicating that there is significant rotation being seen in the line of sight of the disk.  Therefore, the disk is not likely face-on, which agrees with the conclusions made by \citet{shostak89} who calculated the disk's inclination to be near 45\degr.  

Furthermore, the rotation plane of IC 10's main disk would not support the theory that the bridge comes from IC 10's main disk.  To see this, we first note that the spurs morphologically turn inwards \textit{towards} IC 10's main body in a clockwise sense. This is the same sense of winding as the main spiral-like \HI\ features in the disk, suggesting that the whole disk and spurs may rotate and shear clockwise.   If this is the case, then the redshifted motion of the eastern side of IC 10's main disk and blueshifted motion of the western side of the main disk indicate that the southern edge of IC 10's main disk would be the near side of the center and the northern edge of IC 10's main disk would be the far side of the center; otherwise the spurs would not morphologically curve in the correct direction. The northern extension, on the other hand, has to be on the near side pointing towards us if it is a tidal bridge, because tidal bridges move away from their parental galaxy and the northern extension has a blueshifted velocity.  This is opposite of the northern side of the disk, which is pointing away from us.  Tidal features do not generally move into planes that are perpendicular to their parental body, although a bridge connecting two bodies in a close collision could in principle move in any direction.  Therefore, if the northern extension is a bridge, then it is not likely material pulled from the main disk of IC 10.  

The second possibility for the northern extension to be a tidal bridge material is if it is material that has been pulled off of the companion's disk (instead of the disk of IC 10).  This is not likely because there is no obvious tidal arm on the other side of the companion as expected theoretically \citep{toomre72, barnes88, mihos96, mihos04, cox08}.  The lack of a tidal arm is probably not due to the maps not extending far enough north; the combined data maps extend to about -1.3\degr\ in Galactic Longitude and the data from the third project mapping the Magallenic Stream extends much further north than that.  However, no obvious emission associated with IC 10 was found in these regions at the sensitivity of the GBT maps. The northern extension is therefore not likely to be a bridge and companion.

\subsection{An Advanced Merger}\label{merger}
\begin{figure}[!Ht]
\centering
\epsscale{0.45}
\plotone{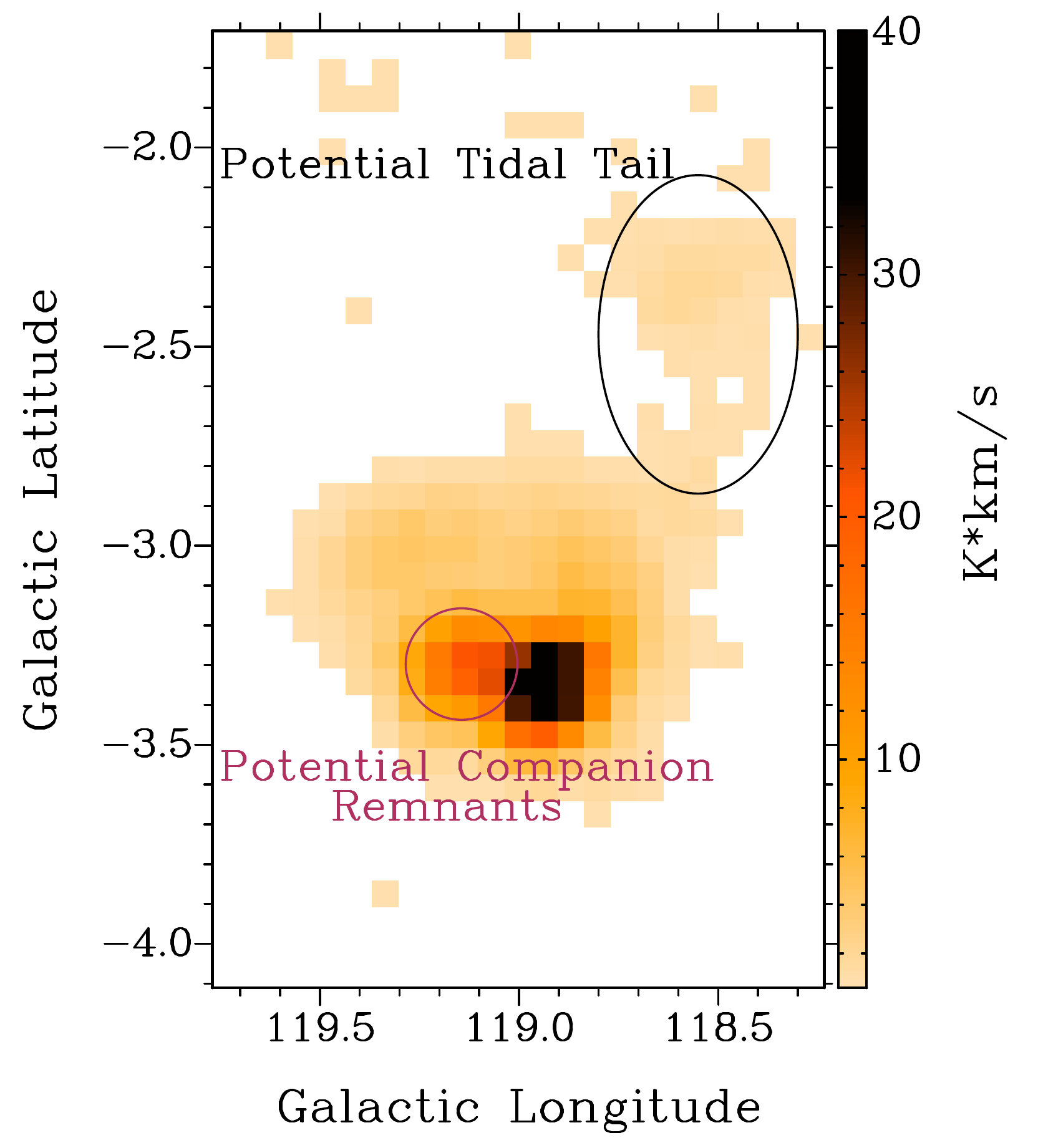}
\caption{A labeled version of Figure~\ref{ic10gbt_ext} showing the location of the features in the case that the northern extension is a tidal tail and the eastern \HI\ peak is a companion. \label{ic10_mergerlabel}}
\end{figure}

\begin{figure}[!Ht]
\centering
\epsscale{0.64}
\plotone{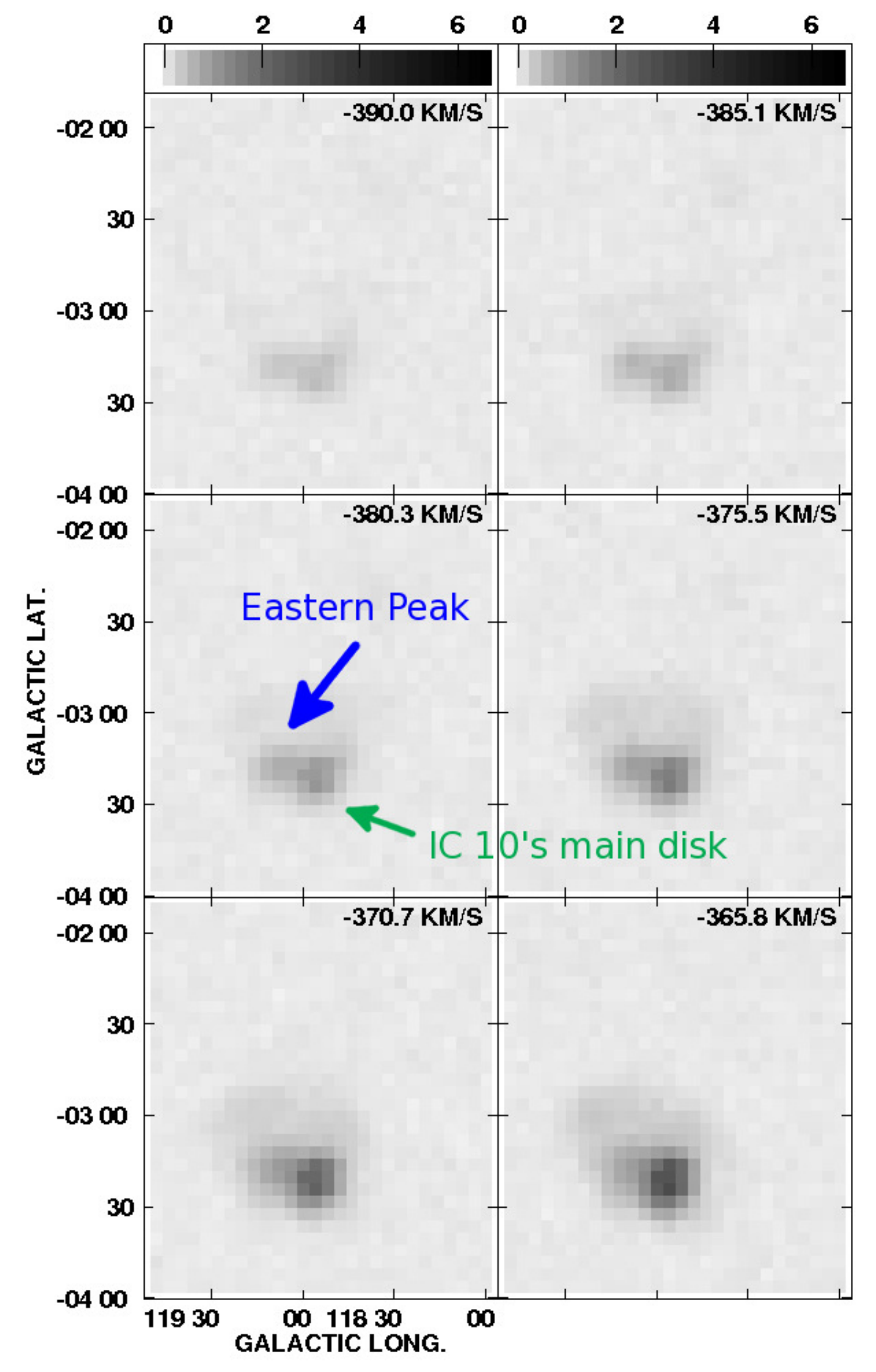}
\caption[IC 10: Channel maps from the GBT data corresponding to the eastern \HI\ peak]{Velocity channels from IC 10's GBT data cube that clearly show the eastern peak. In the velocity range shown, these channels are every third channel in the data cube. \label{ic10_2peaks}}
\end{figure}

A detailed investigation of IC 10's GBT channel maps provides another possible explanation for IC 10's morphology and kinematics: IC 10 may be an advanced merger and the northern extension and southern plume may be tidal tails from the two separate galaxies in that merger (see Figure~\ref{ic10_mergerlabel}).   In Figure~\ref{ic10gbt_ext} there are two \HI\ peaks in the GBT main body of IC 10, separated by a projected distance of \s3 kpc.  This distance was calculated from the distance between the two brightest pixels in the peaks from the \HI\ column density map in Figure~\ref{ic10gbt_ext}.  These two peaks are shown more clearly in Figure~\ref{ic10_2peaks}, where every third channel of the GBT data cube has been plotted from \n389.98 \kms\ to \n365.83 \kms.  The peak to the west is associated with the \HI\ main disk imaged by the VLA.  The peak to the east, however, is not obvious in the VLA map.  It does partially overlap with the eastern spur from the VLA maps in the line of sight.  The eastern peak may therefore be the unresolved eastern spur.  The eastern spur could also be part of the tenuous remnants IC 10's companion galaxy that is merging with IC 10's main disk or part of a tidal feature that was left from a galaxy that has now merged with IC 10's main disk.  The northern extension may then be a tidal tail that has resulted from this merger.

\begin{figure}[!Ht]
\centering
\epsscale{0.48}
\plotone{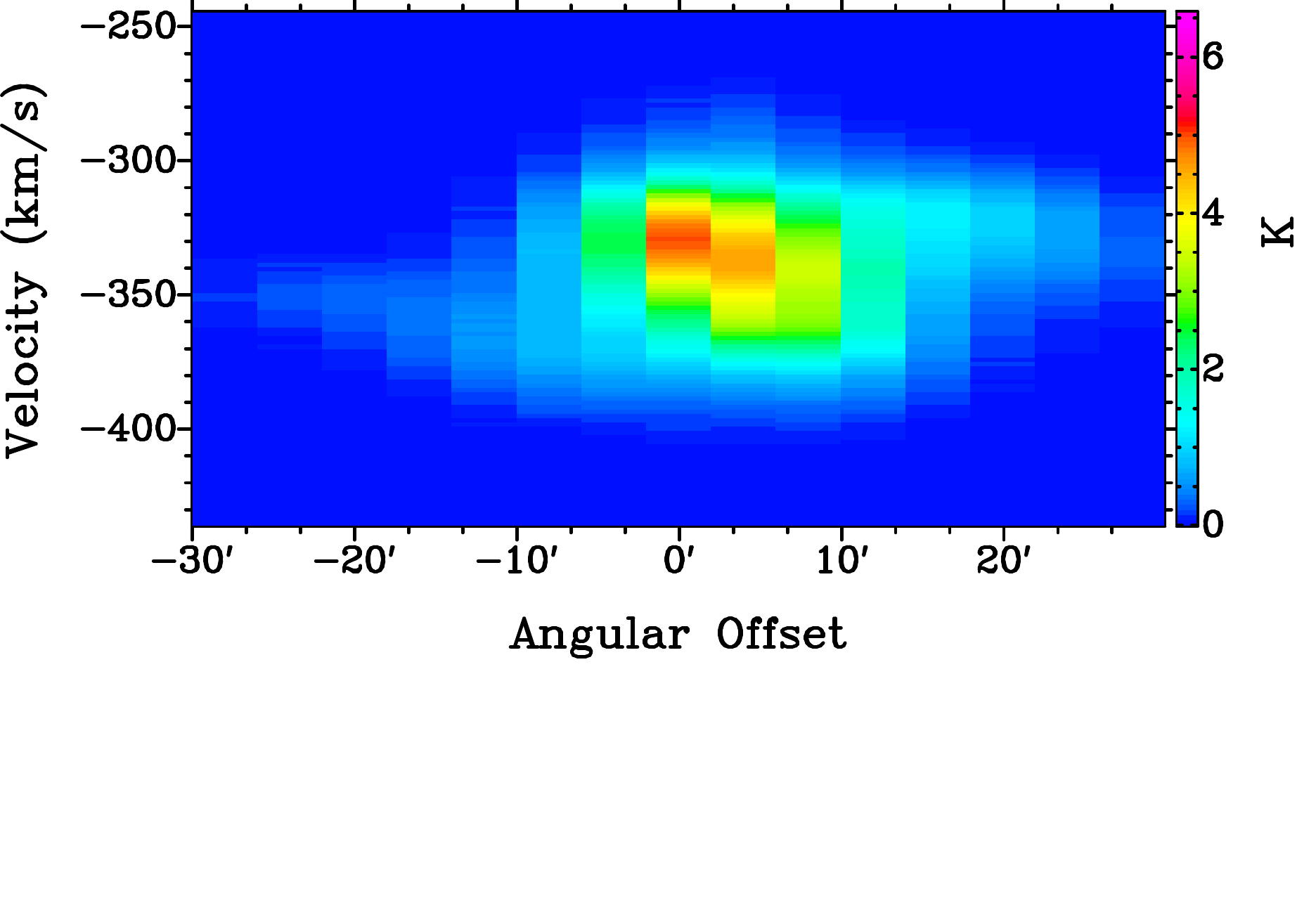}
\plotone{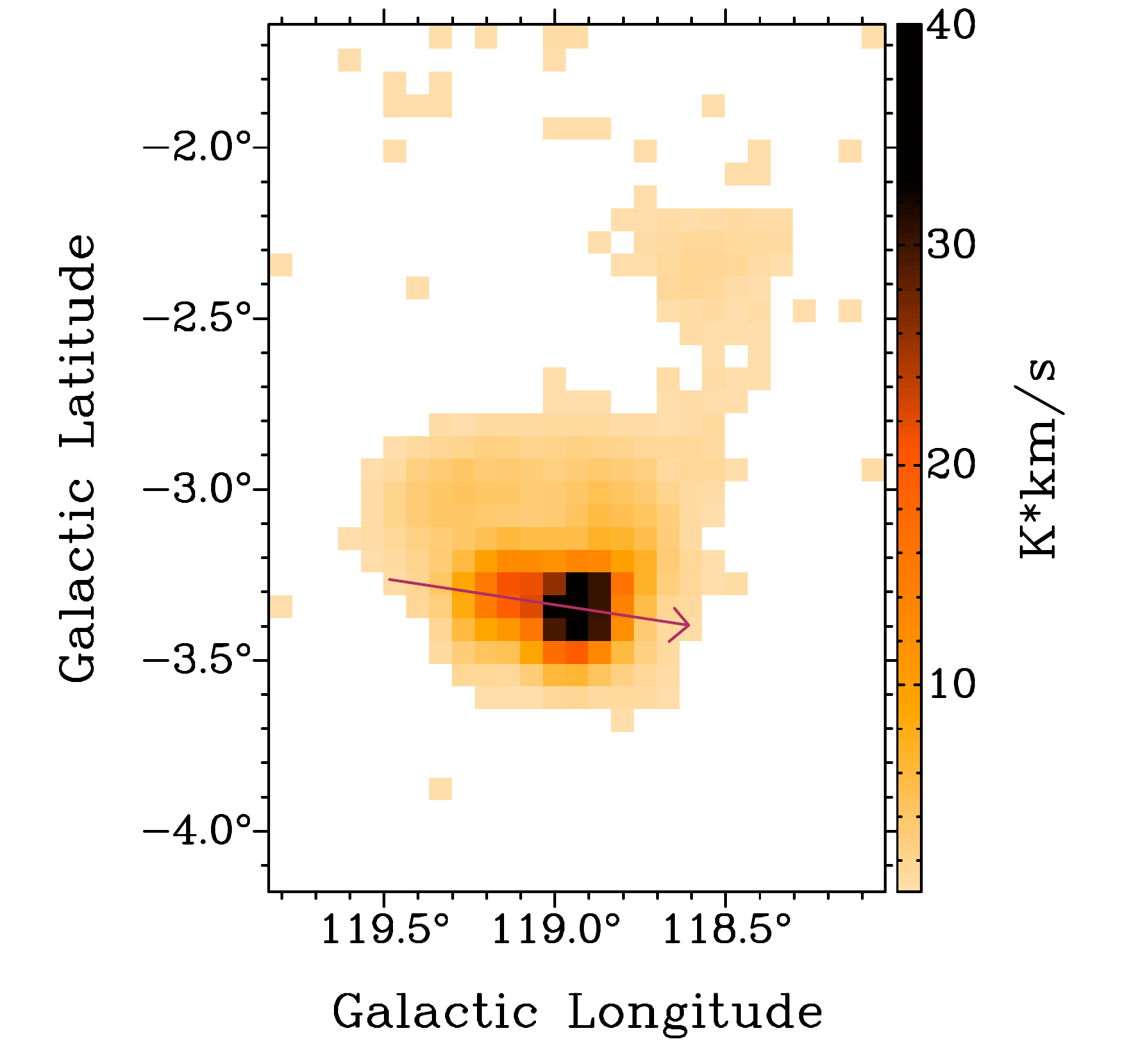}
\caption[IC 10: P-V diagram through both central \HI\ peaks in the GBT data]{IC 10: The left column contains the P-V diagram using the full velocity range of the GBT data cube and the right column contains the GBT integrated \HI\ map from Figure~\ref{ic10gbt_ext} with a red arrow (dark grey in the printed version) indicating the location of the corresponding slice through the galaxy and pointing in the direction of positive offset. This slice goes through both central peaks seen in the GBT data.  \label{ic10pv_2peaks}}
\end{figure}

To see if the two peaks are kinematically distinct, a P-V diagram with a slice going through both peaks was made with the full GBT data cube and is shown in Figure~\ref{ic10pv_2peaks}.  The western peak is the brightest emission in the P-V diagram. The overall velocity trend in the western peak is the same as that seen in the VLA data: decreasing velocities from left to right.  The eastern peak is the faint emission at \n390 \kms\ to \n350 \kms\ and \n1.5\arcmin\ to \n13\arcmin.   The eastern peak shows up below the left side of the western peak at lower velocities.  This difference in the P-V diagram indicates that the eastern peak is not rotating in a solid body manner with the western peak.  

To investigate if the kinematics of the rest of the GBT \HI\ data support this idea, a new velocity map was made to exclude the western \HI\ peak and the southern plume of IC 10.  This was done in two steps: first the data cube was blanked by hand in AIPS using \textsc{blank} to remove as much of the western peak as possible and second, a velocity range of \n422.2 \kms\ to \n330.4 \kms\ was used to exclude the southern plume when integrating the channels to make a new map of the GBT \HI\ data.  The first \textsc{blank}ing step was done by visual inspection of individual velocity channel maps; since the gas associated with the western peak has a distinct velocity range in the data cube (see Figure~\ref{ic10pv_2peaks}) and has a high flux, much of its emission was distinguishable from other emission in the individual velocity channel maps.

\begin{figure}[!Ht]
\centering
\epsscale{0.55}
\plotone{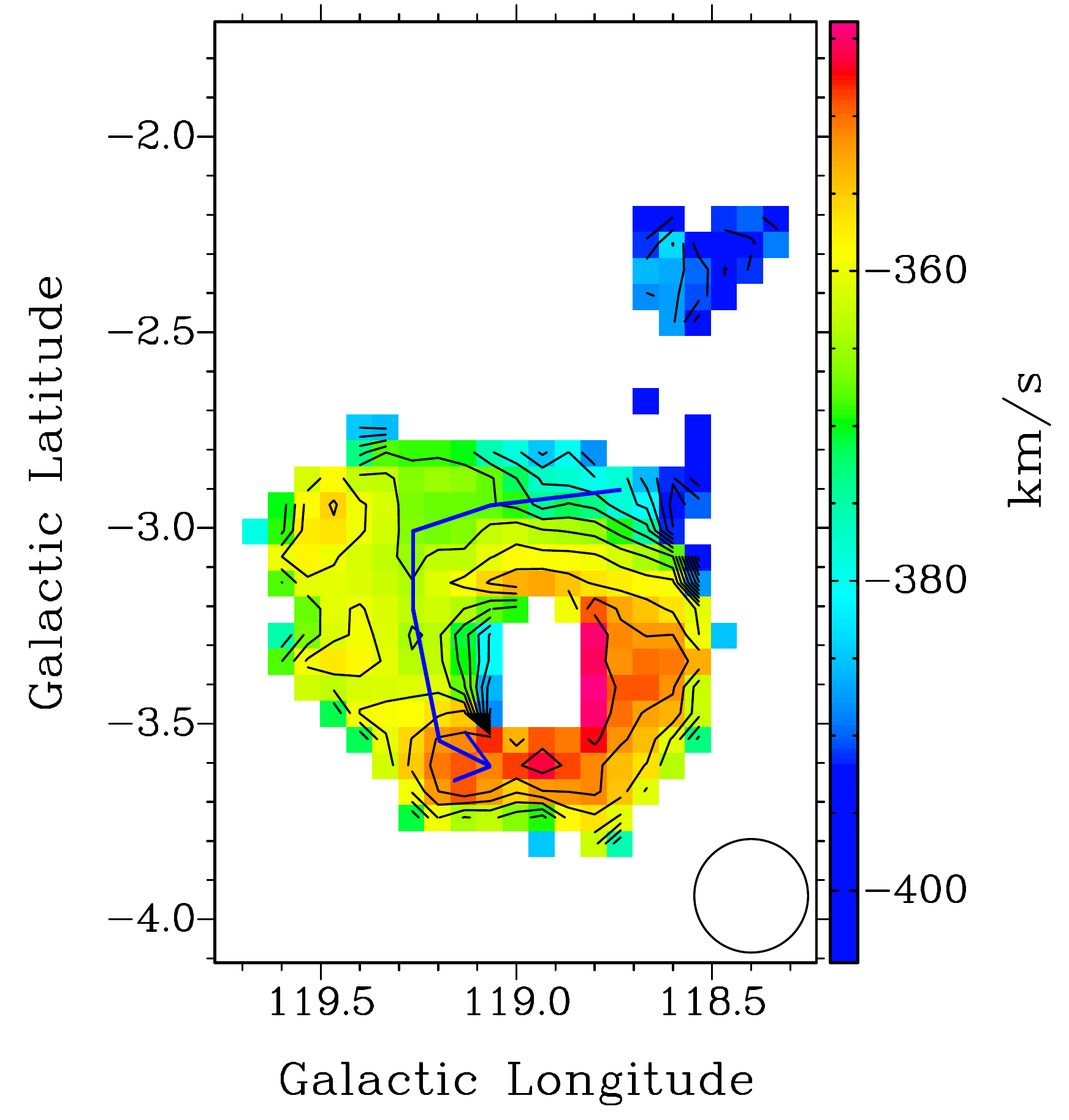}
\caption[IC 10: GBT velocity field without the southern plume and the western peak]{Velocity field of IC 10's GBT data without the southern plume or most of the western peak.  The velocity contours are from -404 to -348 \kms\ in intervals of 4 \kms.  The blue line (dark grey in the print version) represents the possible transition of the northern extension into the main body of IC 10. \label{ic10_nombnopl}}
\end{figure}

The resulting velocity field is shown in Figure \ref{ic10_nombnopl}.  From this figure we see that the line of sight velocities from the northern extension transition into the main body from the northwest to the northeast (as in Figure~\ref{ic10gbt}c and Figure~\ref{ic10_davidsvideo}), and then spiral counterclockwise in towards the center of the galaxy (traced by the blue line).  Much of this velocity field could be tracing a wake or tidal feature spiraling in towards one of the two central peaks.  Depending on the properties of the interaction and the two galaxies interacting, a tidal feature can appear to be spiraling as a merger occurs.  For example, in \citet{cox08}, their Figure 4 shows the resulting time steps of a merger between two galaxies with a total mass ratio of 2.3:1 resulting in many spiraling tidal features in the gas component.  The lower mass galaxy also quickly merges with the larger mass galaxy prior to the tidal tails dissipating.  It is therefore possible that IC 10 has merged with a companion of unequal mass.

\begin{figure}[!Ht]
\centering
\epsscale{0.6}
\plotone{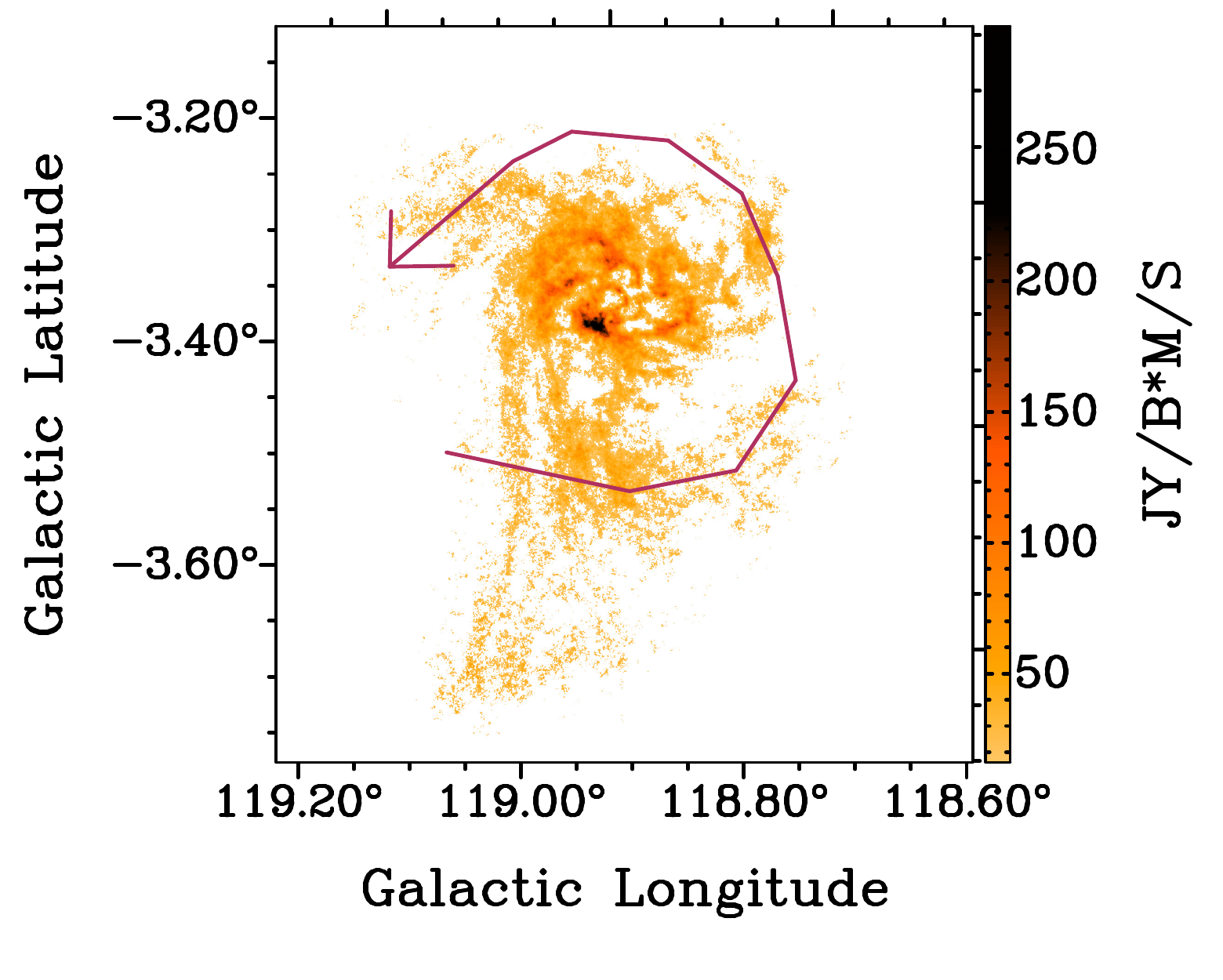}
\caption{The VLA \HI\ column density map from Figure~\ref{ic10vla} and labeled with a red arrow (dark grey in the printed version) showing the morphological counterclockwise direction of the spurs around IC 10's main disk.\label{ic10countercw_spurs}}
\end{figure}

If IC 10 is the result of a merger, then it is possible that the spurs in IC 10 may also be related to the tidal disturbance in Figure~\ref{ic10_nombnopl}. As noted in section~\ref{bridge_explanation}, the eastern spur overlaps with the arc of material that appears to be the northern extension material transitioning into IC 10's main body (see the white arrow in Figure~\ref{ic10gbt}c).  Figure~\ref{ic10_nombnopl} shows that the arc from Figure~\ref{ic10gbt}c may continue into the main body towards the southwestern spur as indicated by the blue arrow.  Together the spurs also appear to point morphologically in a counterclockwise direction \textit{around} IC 10's main disk towards the eastern peak as shown in Figure~\ref{ic10countercw_spurs}.  The spurs also have velocity gradients opposite of IC 10's VLA main disk, and are therefore unlikely to have come from it.  These spurs may all be part of a tidal disturbance that wraps around IC 10's main disk and appear as spurs in the IC 10 VLA maps due either to density enhancements or line of sight overlap with tenuous emission from the edge of IC 10's main disk.  Also, in this scenario, since the spurs have not originated from IC 10's main disk, the tidal feature that they are a part of would have come from IC 10's companion with which IC 10 has merged.  

\begin{figure}[!Ht]
\centering
\epsscale{1}
\plotone{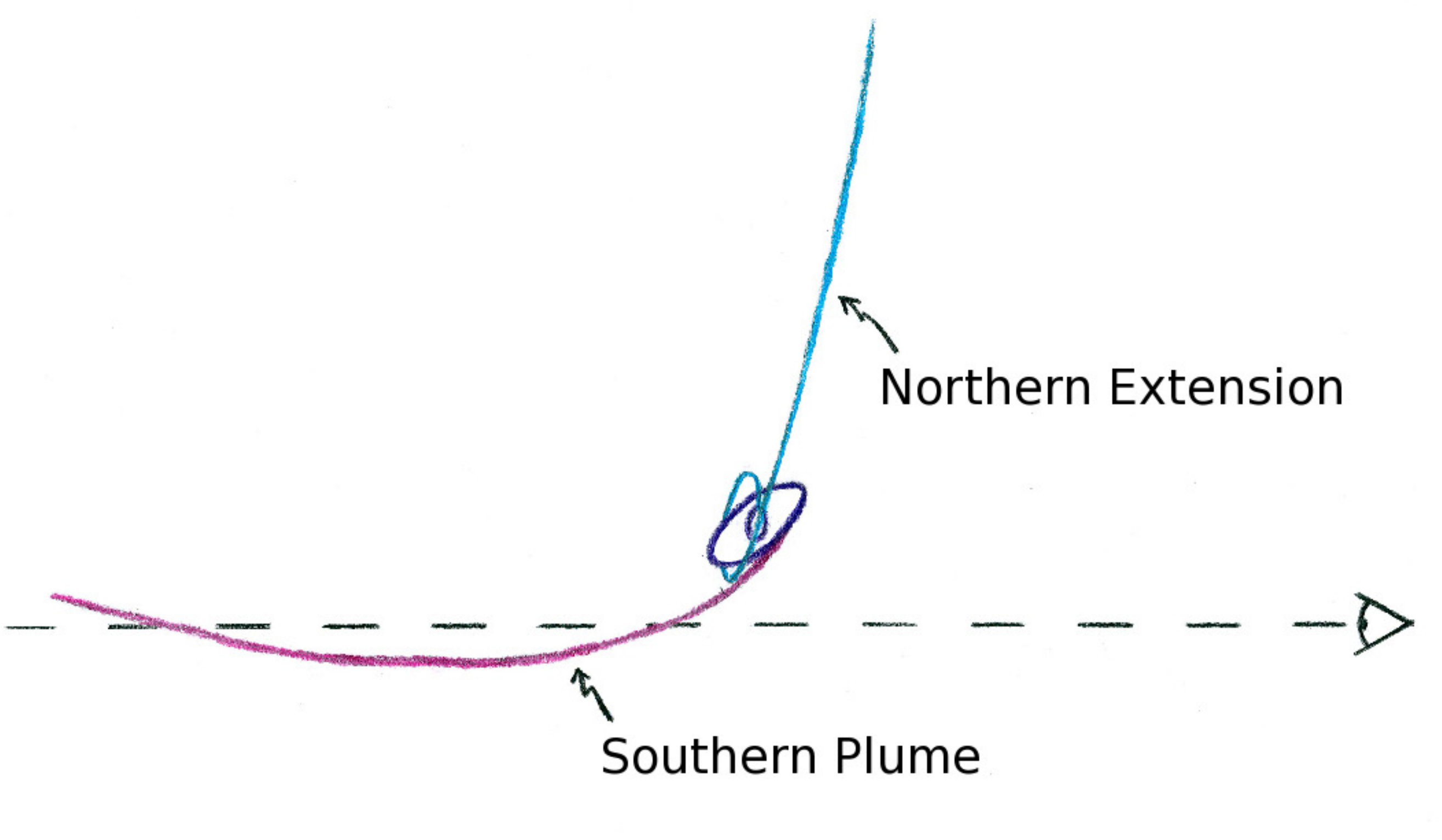}
\caption{A simple schematic of a potential viewing angle of IC 10 for the advanced merger model.\label{ic10_schematic}}
\end{figure}

\begin{figure}[!Ht]
\centering
\epsscale{0.7}
\plotone{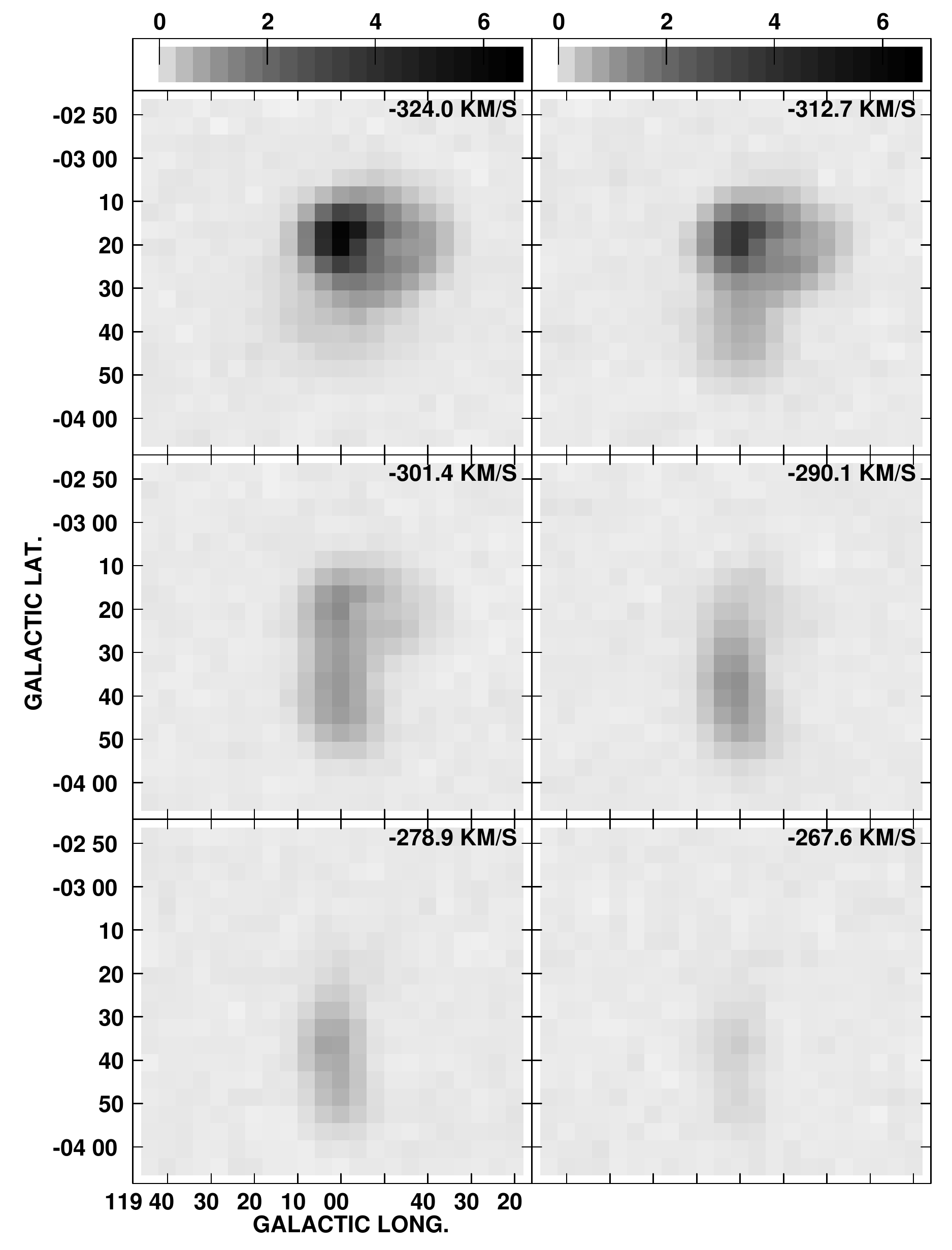}
\caption{Velocity channel maps for IC 10's southern plume at intervals of 7 channels.\label{ic10_southernplume}}
\end{figure}

The southern plume could also be a second tidal tail seen at an extreme projection angle.  A schematic is shown in Figure~\ref{ic10_schematic} to demonstrate a possible viewing angle of the configuration.  If the southern plume is a tidal tail that is extending along the line of sight, then it will also likely have a velocity gradient.  In Figure~\ref{ic10_southernplume}, every seventh velocity channel containing obvious southern plume emission from the GBT data cube has been plotted.  The channels shown here cover a large amount of velocity space: 56 \kms. The southern plume grows in vertical length in the first three channels plotted, and then becomes shorter in length in the last three channels.  It also has a lower column density in the last two plotted channels.  These features are consistent with an elongated feature extending along the line of sight.  As seen in the VLA velocity map in Figure~\ref{ic10vla}b, the velocities of the southern plume transition well into the main disk of IC 10; as the southern plume gets closer to the main disk in IC 10's VLA velocity map, its velocities go from \n280 \kms\ to about \n300 \kms\ and it reaches the velocities of the redshifted, east side of IC 10's main disk as it meets it in the line of sight.  Therefore, if the southern plume is a tidal tail, it is likely to have originated from IC 10's VLA main disk (the western peak in the GBT map; see Figure~\ref{ic10_2peaks}).   

Tidal tails are created from material that has been pulled off of the outer edge of a galaxy \citep{toomre72} and thus should retain some of the motion of the edge of the disk.  Since the southern plume is redshifted with respect to IC 10's systemic velocity, it would be material that was pulled out of the disk while moving away from us with respect to the center of the disk, putting the tidal tail on the far side of the center of the disk. The southern plume is not only redshifted with respect to IC 10's systemic velocity (see Figures~\ref{ic10vla}b and \ref{ic10gbt}c), but it also extends to the south of IC 10's main disk.  This indicates that the redshifted eastern edge of IC 10's main disk is likely moving towards the south in the line of sight, otherwise the southern plume would not be to the south of the disk.  The downward motion combined with the redshift of the eastern edge of IC 10's main disk and blueshifted western edge of IC 10's main disk indicates that the disk has to rotate counterclockwise, with the northern edge of the disk on the near side and the southern edge on the far side.

The disk orientation required for this merger scenario is the opposite of that mentioned in section~\ref{bridge_explanation}.  In section~\ref{bridge_explanation}, the orientation of the disk was reasoned under the assumption that the spurs originated from IC 10's main disk and were therefore following a clockwise rotation and shear pattern of the disk.  The spurs in this merger scenario, however, are part of a tidal feature that has originated from the companion with which IC 10 has merged and thus do not need to follow the rotation/shear pattern of IC 10's    main disk. However, without the spurs giving an indication of the rotation pattern in IC 10's main disk, it is difficult to tell which direction the disk is rotating.   When IC 10's main disk is viewed alone in the VLA maps (without the spurs or southern plume), the densest \HI\ appears to be spiraling in a somewhat clockwise direction, which could be indicating shear in the disk due to clockwise rotation.  If that is the case, then the southern plume cannot be a tidal tail that was pulled from IC 10's main disk.  However, there are a lot of holes in the main disk of IC 10, and these holes could be creating the \textit{illusion} of shear in the disk.  

\begin{figure}[!Ht]
\centering
\epsscale{0.8}
\plotone{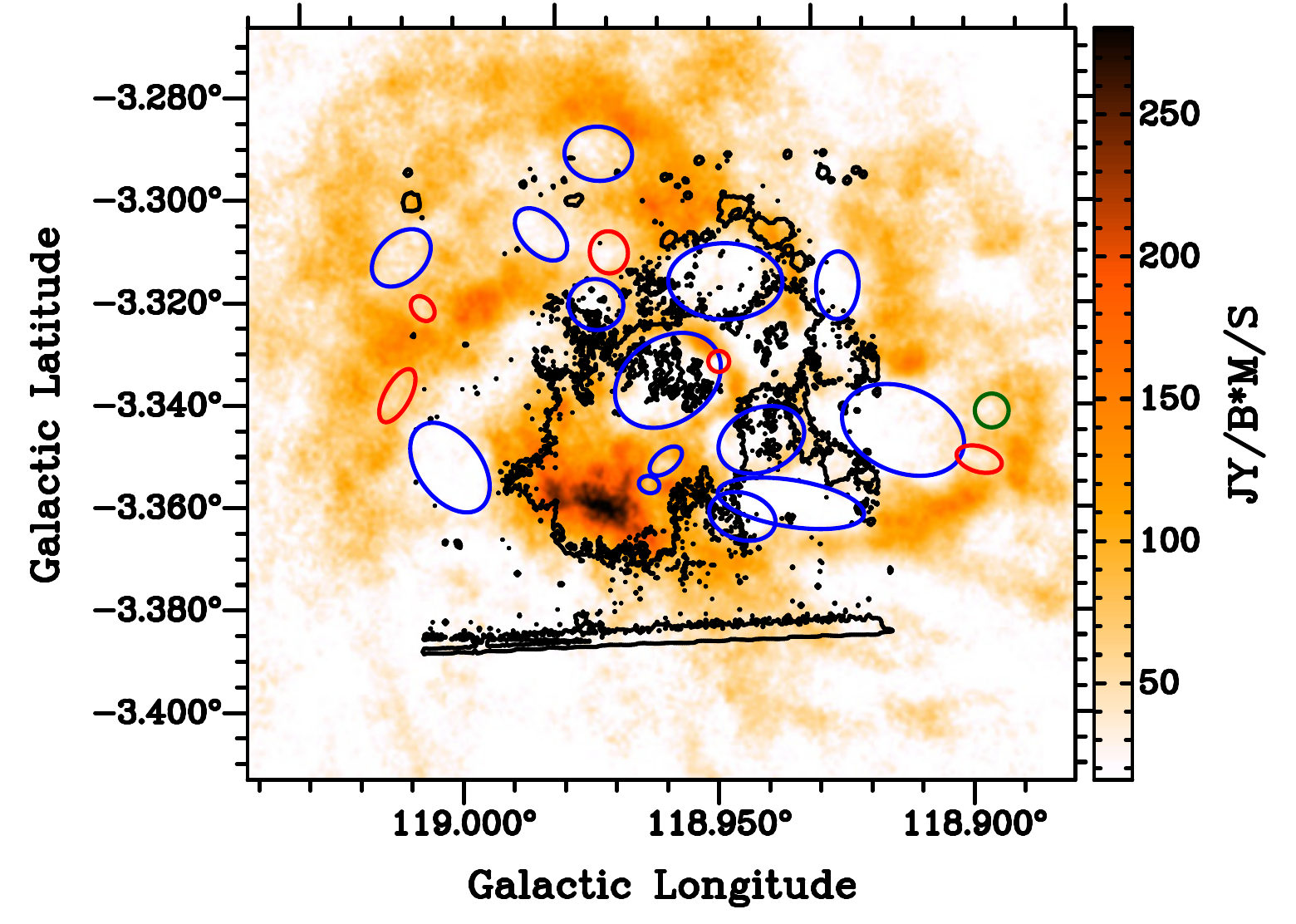}
\caption{The main \HI\ disk of IC 10 from the VLA maps.  The circles indicate the location and morphology of holes located in IC 10's \HI.  The blue holes (for color, see the online version) are empty cavities that have no detectable HI along the projected line of sight, the red holes have HI either on the near or far side of the cavity, and the green hole has HI detected on both the near and far sides but is empty in the center  The black contours are the H$\alpha$ emission (see Figures~\ref{ic10_star} and \ref{ic10starkntr} for images of the H$\alpha$ emission alone and the H$\alpha$ contours with only the VLA \HI\ as a colorscale, respectively).  \label{ic10_holes}}
\end{figure}

To investigate this further we plot the location of holes identified in IC 10 in Figure~\ref{ic10_holes}.  These holes were identified as part of the LITTLE THINGS holes and shells analysis (Pohkrel, \et\ in preparation).  The holes were detected using both natural and robust weighted integrated HI VLA maps.  Potential holes and their sizes were first identified using the task \textsc{kvis} in the visualization software Karma\footnote{Documentation is located at  \url{http://www.atnf.csiro.au/computing/software/karma/}.} \citep{gooch96}.  The quality of each hole was then determined using the same methods as those in \citet{bagetakos11}.  The hole quality was determined by verification of its existence in at least three sequential channel maps and examination of position-velocity (P-V) diagrams in the task \textsc{kpvslice}.  For IC 10, 20 high quality holes were identified (plotted in Figure~\ref{ic10_holes}), with diameters of 46 pc to 250 pc.  This is more than twice the number of holes discovered by \citet{wilcots98}; we confirm seven of the eight holes identified in \citet{wilcots98} and we also identify a more symmetric distribution of holes in the plane of the sky than \citet{wilcots98}.  Each hole also has an assigned type (indicated by their color in Figure~\ref{ic10_holes}) based on line of sight velocities in P-V diagrams: blue holes are empty cavities that have no detectable HI along the projected line of sight, red holes have HI either on the near or far side of the cavity, and green hole has HI detected on both the near and far sides but is empty in the center.  The location of the edges of the holes in Figure~\ref{ic10_holes} match up well with most of the dense \HI\ that gives IC 10's main disk the spiraling \HI\ features.  This could indicate that the \HI\ shells surrounding the holes are creating a spiraling illusion in the main disk.  For the holes that have at least one side (near or far) intact, an upper limit of the kinetic age of each hole was calculated using the expansion velocity of the hole in the line of sight and the size of the hole.  The age range calculated for IC 10's holes is 1 Myr to 7.6 Myr.  This indicates that the red and green holes are quite young and the stellar components that made them would likely still be visible in the H$\alpha$ emission since it probes star formation over the past 10 Myr.  Unfortunately we cannot compare all of the holes to the H$\alpha$ emission because the map of IC 10's H$\alpha$ emission is limited to a box with a bottom left corner at about (119\degr, \n3.38\degr) and an upper right corner of about (118.94\degr, \n3.27\degr) as can be seen in the left side of Figure~\ref{ic10_star}.  However, not all of the holes within the limits of the H$\alpha$ map have H$\alpha$ associated with them.  This may be because none of these holes in this region have both sides of their shells intact, that means that their ages are highly uncertain, and they could be quite old.  In fact, the only hole with both sides intact is outside the limit of the H$\alpha$ map.  So it is possible that these holes are older and are creating a spiraling shape in the \HI\ which imitates clockwise shear, which means that it is possible that the disk may be rotating in a counterclockwise direction.

The two tidal tails (the southern plume and the northern extension) thus appear on opposite ends of IC 10 and are symmetric in velocity (the first is redshifted while the second is blueshifted).  The masses of the two potential tidal tails, however, are vastly different.  A rough mass estimate was made from the GBT data cube where the western peak has been blanked: the velocity channels that contain the southern plume were integrated to get a rough estimate of the box size needed to contain the emission and then \textsc{ispec} was used to get a total flux in the channels containing the southern plume emission.  The resulting mass of the southern plume is \s10$^{7}$\ $M_{\sun}$ or 10\% of the total GBT mass.  The mass of the northern extension was measured to be $6\times10^{5}\ M_{\sun}$.  Such a large difference in mass indicates that the two tidal tails were likely made by the two different galaxies merging.  This would make IC 10 similar to the antennae (NGC 4038/NGC 4039) and mice galaxies (NGC 4676) where each galaxy has produced a tidal tail, but at a different viewing angle.

\citet{nidever13} showed that the northern extension was not due to star formation pushing the gas outwards, but is it possible that the southern plume could be material blown out from stellar feedback?  The southern plume is quite massive at \s10$^{7}$\ $M_{\sun}$, so if it is blow out from star formation, it would be a lot of material that has left the disk.  The plume extends \s7 kpc in the GBT maps, and IC 10 has a starburst about 10 Myr old, so if the current starburst formed the southern plume with stellar feedback, the outflow velocity required would be \s700 \kms.  This would be a very high outflow speed, therefore, it is unlikely that IC 10's southern plume was created from outflow that came from the current starburst.  

The change in velocity across the top round part of the northern extension can also be explained through a tidal tail.  The velocity field at the end of the northern extension, centered around (118.5\degr, \n2.3\degr), shows a turnaround in velocity back towards less negative velocities, from about \n395 \kms\ to \n385 \kms.   The systemic velocity of IC 10 is \n348 \kms.  The velocities at the end of the northern extension could be indicating that the gas at the end of the tidal tail is beginning to fall back towards the galaxy.

The \HI\ velocity dispersion map as measured by the VLA (Figure~\ref{ic10vla}c) fits well with the theory of IC 10 being an advanced merger.  There are generally high velocity dispersions of \s20 \kms\ throughout the main disk of IC 10, indicating that the galaxy is quite disturbed as would be expected for an advanced merger.  It is also possible that the high velocity dispersions in IC 10's main disk are in part or fully due to the starburst in IC 10.  The southern plume also has significantly higher velocity dispersions (20-30 \kms) where, as seen in the schematic in Figure~\ref{ic10_schematic}, the southwestern spur would likely be crossing the southern plume in the line of sight.  This jump in velocity dispersion would be expected for two features of different velocity ranges crossing each other in the line of sight.  The southern plume then has lower velocity dispersions as it gets further from the main body of IC 10.  The spurs have this same pattern of low velocity dispersions further from the main body.  Low velocity dispersions are not unusual for tidal tails \citep{hibbard94, hibbard96, hibbard99, mihos01}.  Therefore, the higher velocity dispersions in the spurs closer to the disk could be caused by the gas in the spurs closer to the main body overlapping with some of the extended emission from IC 10's main disk. 

The stellar kinematics of IC 10 were observed by \citet{gonclaves12} by measuring the motion of planetary nebulae (PNe) with ages of \s8 Gyr.  They concluded that the kinematics of these PNe followed roughly the kinematics of the VLA \HI\ maps from \citet{wilcots98}, including the western spur.  Some of these PNe may therefore be a part of the tidal feature where the PNe are stars that have been pulled from the outskirts of the potential companion with which IC 10 has merged.  The only stellar components found in DSS and NED near the eastern peak were labeled as part of IC 10.  However, no obvious stellar disk was seen in the location of the eastern peak in DSS.  Deeper stellar images of this region are needed to confirm if there is a stellar component to the eastern peak or other emission associated with the potential tidal features.

\subsection{A Galaxy Still in Formation}\label{infalling}

\begin{figure}[!Ht]
\centering
\epsscale{0.45}
\plotone{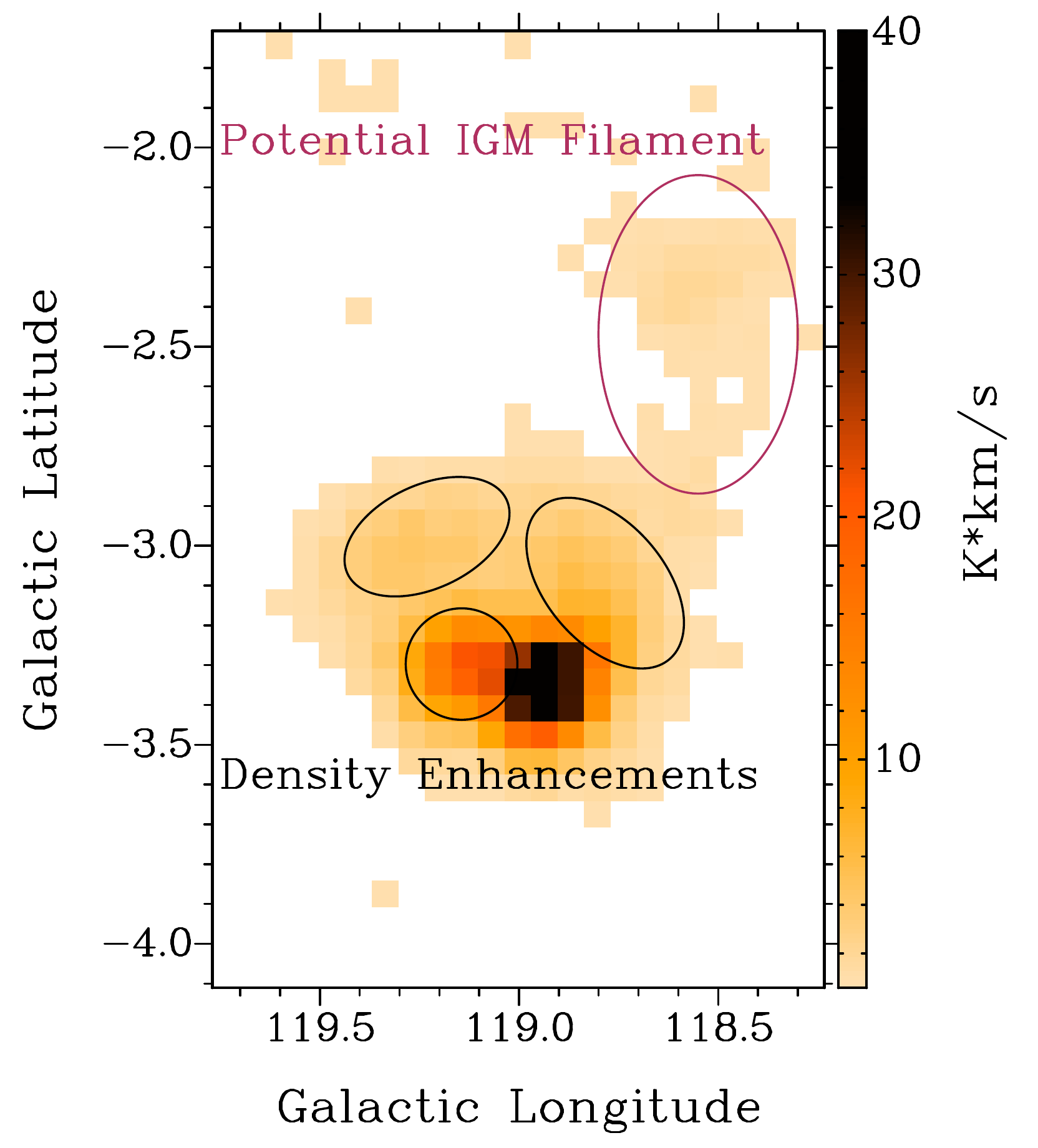}
\caption{A labeled version of Figure~\ref{ic10gbt_ext} showing the locations of the column density enhancements and the potential filament of IGM. \label{ic10_filamentlabel}}
\end{figure}

\citet{shostak89} and \citet{wilcots98} suggested that IC 10 may be a galaxy still in the process of forming by passively accreting the surrounding IGM.  Since the northern extension extends 7 kpc from the disk and is blueshifted, it could be gas on the far side of IC 10 that is being accreted onto IC 10 (see Figure~\ref{ic10_filamentlabel}).  \citet{keres05} model accretion onto low mass galaxies and find that the main mode of accretion is cold accretion through filaments.  \citet{brooks09} also find that cold accretion is a dominant form of IGM accretion in low mass galaxies and suggests that the filaments will accrete onto the galaxy essentially in free fall. In this case, gas that is passively accreted onto a galaxy from the IGM would be expected to have velocities that increase as the gas gets closer to the disk.

Energy conservation was used to estimate how much the velocity would increase along the length of the northern extension.  A point mass potential was assumed for the gravitational potential well, resulting in the following equation for the change in velocity:
\begin{equation}
v_{f}=\sqrt{2GM\left(\dfrac{1}{r_{f}}-\dfrac{1}{r_{i}}\right)}
\end{equation}
where $r_{i}$ and $r_{f}$ are the initial and final radii respectively and \textit{M} is the mass of IC 10.  The mass of IC 10's disk was taken to be the dynamical mass determined from mass models of the LITTLE THINGS VLA data: 2.6$\times$10$^8$ $M_{\sun}$ (Oh, \et\ in preparation).  The calculated velocity change along a filament falling into a galaxy like IC 10 from cold accretion is then \s10 \kms.  This is a small change in velocity for the 7 kpc length of the northern extension.  The northern extension is also very elongated to the north of IC 10's main body.  It is therefore likely that if it is a filament being accreted onto IC 10, then most of the velocity increase will be perpendicular to the line of sight.  This means that we may not see the increase in velocity along the filament because motion in the plane of the sky does not produce a Doppler shift.  It is therefore possible that the northern extension is an intergalactic gas filament being accreted onto the far side of IC 10.

\begin{figure}[!Ht]
\centering
\epsscale{0.8}
\plotone{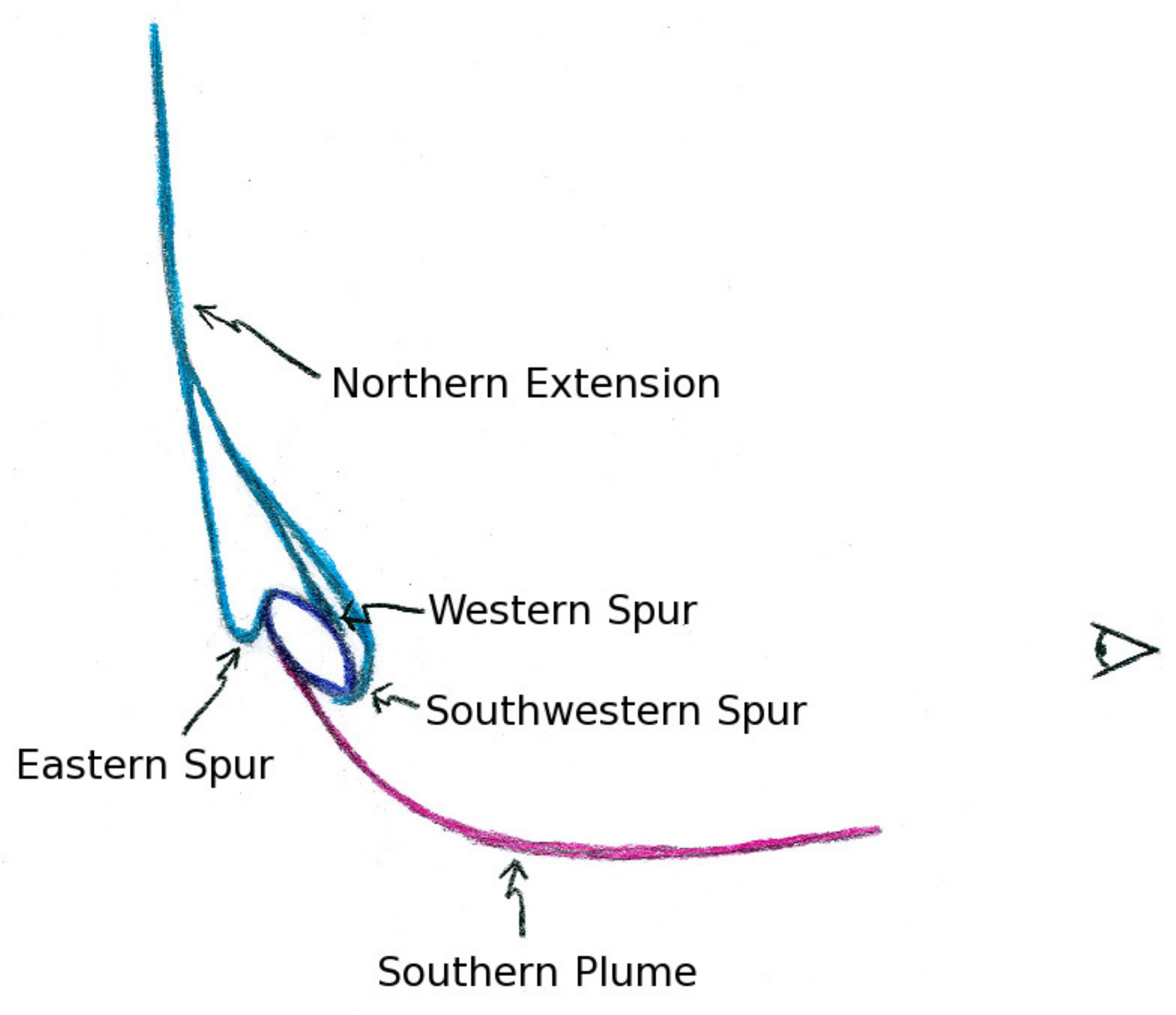}
\caption{A simple schematic of a potential viewing angle of IC 10 for the filament accretion model.\label{schematic_filament}}
\end{figure}

The southern plume has a distinct velocity range which points to the southern plume being a distinct feature.  If the northern extension is an infalling gas filament, then perhaps the southern plume is also an infalling IGM filament feeding IC 10's main disk.  The mass of the southern plume is 10$^{7}$\ $M_{\sun}$ or 10\% of the total GBT mass.  \citet{sancisi08} discuss external gas clouds surrounding galaxies and find many examples of gas clouds with masses of \s10$^{7}$\ $M_{\sun}$.  They suggest that some of these gas clouds and some of the high velocity gas clouds surrounding the Milky Way likely have an IGM origin.  Therefore, the southern plume has a reasonable mass for an accreting IGM filament.  The southern plume has emission in channels extending over a total of \s56 \kms\ as seen in Figure~\ref{ic10_southernplume}; and in the integrated GBT maps, a velocity extent totaling 35-40 \kms\ as seen in Figure~\ref{ic10gbt}c.  If we assume that the southern plume extends to about the same length as the northern extension's projected length, then the change in velocity along the filament only needs to be \s10 \kms\ if it is in free fall.   The southern plume also consistently becomes more redshifted (relative to the systemic velocity of IC 10) as it gets further away from the disk. This velocity trend is most obvious in Figure~\ref{ic10gbt}c where the southern plume's velocities become more redshifted as the southern plume gets further from the main disk of IC 10 in radius.  If the southern plume is a filament accreting onto IC 10 from the nearside of the disk, then it seems unlikely that it would be moving slower as it gets closer to the disk in the line of sight.   One possible explanation for this velocity trend is that the southern plume is bent in such a way that it is aligned more perpendicular to the line of sight as it gets closer to the disk and is aligned more with the line of sight as it gets further from the disk (see Figure~\ref{schematic_filament}).  Therefore, we would be seeing a larger component of the total in-streaming velocity caused by the infall of the filament in the most southern part of the southern plume.  Then, the large decrease in southern plume's velocity as it approaches IC 10's main disk may not be large at all, rather, just some component of the southern plume's velocity is lost to the plane perpendicular to the line of sight as it gets closer to the disk.  Without initial conditions of the accretion (such as relative speed), it is then difficult to determine the actual change in velocity along the filament.

The general velocity trends of the northern extension and the southern plume are sensible if the gas is accreting.  The northern extension has blueshifted velocities relative to IC 10's systemic velocity; if it is accreting onto IC 10, then it must be coming from the background moving towards the disk of IC 10.  The southern plume is redshifted, and therefore the gas would be accreting onto the main body of IC 10 from the near side of the disk.  Then the northern extension would be on the far side like the northern side of the main disk (see Section~\ref{bridge_explanation}) and the southern plume would be on the near side like the southern side of the main disk.  The accretion could therefore be feeding the disk in the same plane as shown in the schematic in Figure~\ref{schematic_filament}.

If the northern extension observed with the GBT is an accreting filament, then can it explain some of the unusual features seen in the VLA map of the main disk? IC 10's main disk seen in the VLA maps is not likely the parental disk to the \HI\ spurs (Figure~\ref{ic10vla}) since the spurs have velocity gradients opposite to that of the main disk.  The spurs could instead be part of the accreting filament.  In the \HI\ column density map in Figure~\ref{ic10gbt_ext}a there are several \HI\ density enhancements seen in the data: one to the north of the main VLA disk, one in the northeast of the GBT's main body, and the GBT's eastern peak (see Figure~\ref{ic10_filamentlabel}).  These density enhancements could be turning points in the line of sight in the filament as it feeds gas to IC 10's disk.  These density enhancements also line up well with the spurs in the VLA maps.  The northern extension could then be a filament that is splitting and feeding the disk through all three spurs.

The VLA velocity dispersions in Figure~\ref{ic10vla}c also support this interpretation.  The eastern spur is blueshifted with respect to the systemic velocity of the galaxy, but it is meeting the redshifted western side of IC 10's main disk in the VLA data.  At this juncture, there is gas with dispersions of \s30-40 \kms.  Although these higher dispersions do occur in the spur along the line of sight, they are close to the disk and may be a result of the filament gas hitting the disk in a retrograde fashion. The southwestern spur also exhibits this type of behavior.  The part of the spur that is furthest from the disk in projection is redshifted, but as the southwestern spur gets closer to the disk, its velocity nears the systemic velocity of the galaxy (the green part of the southwestern spur in Figure~\ref{ic10vla}b).  Therefore, the part of the southwestern spur making contact with the disk is moving at approximately the systemic velocity of IC 10, while the disk at the same location is more redshifted.  This location also has high velocity dispersions of \s30-40 \kms, indicating a possible overlap of two gas components on the line of sight or a collision-caused increase in turbulence.  The western spur does not fit as simply into this picture: it is redshifted and hitting the blueshifted side of the disk relative to the systemic velocity.  Although the velocity dispersions are high at this juncture, \s20 \kms, the entire main disk of IC 10 in the VLA data has high velocity dispersions near \s20 \kms.  Perhaps the western spur does not hit the disk directly on the western edge of the disk and the higher dispersions near (118.94\degr, \n3.32\degr) are from the western spur hitting the disk closer to the center of the disk.  The outer edges of the spurs (the parts furthest from the main body of IC 10) all also have lower velocity dispersions.  This is consistent with them being inflowing cold gas.

We have estimated the rate of accretion as $(M\times V)/L$, where \textit{M} is the mass of the gas being accreted onto IC 10, \textit{V} is the velocity of the gas relative to the disk of IC 10, and \textit{L} is the length of the gas filament.  For the northern extension, $M=6$$\times$10$^5$\ $M_{\sun}$, $V=15$ \kms\ (from Figure~\ref{ic10gbt_ext}), and $L=7$ kpc, resulting in an accretion rate of 0.001\ $M_{\sun}$\ yr$^{-1}$.  The northern extension will then finish accreting onto IC 10 in \s0.6 Gyr. If we assume that the southern plume is an accreting filament and also the same length as the northern extension, then using $V=30$ \kms\ and $M=10$$^{7}$\ $M_{\sun}$ we get an accretion rate of 0.05\ $M_{\sun}$\ yr$^{-1}$.  With these assumptions, the southern plume will fully accrete onto IC 10 in \s0.2 Gyr.  From Table~\ref{tab:galinfo} we note that the star formation rate for IC 10 is \s0.08\ $M_{\sun}$\ yr$^{-1}$\ kpc$^{-2}$.  The accretion rate estimate of the northern extension is not able to account for this star formation rate, but the accretion rate estimate of the southern plume is close to this.  The accretion from the southern plume could therefore be fueling the starburst in IC 10.  Models described in \citet{verbeke14} also suggest that the southern plume would likely trigger a strong starburst in IC 10.   \citet{verbeke14} model the formation of BCDs through accretion of IGM onto a dwarf irregular galaxy.  In their models, they find that gas clouds $\ge$10$^{7}$\ $M_{\sun}$ are necessary to trigger a starburst.  The southern plume fits this criterion.  

Two filaments potentially feeding IC 10's disk could explain what has fueled IC 10's burst of star formation. The presence of two filaments around IC 10, with velocities suggestive of accretion from both sides, is consistent with current models of galaxy formation at the nodes of cosmic gas filaments \citep[e.g.,][]{dekel09}.  Of the two potential filaments, the southern plume is the one that would be most likely to trigger a burst in IC 10 because it is massive enough to do so according to models described in \citet{verbeke14}. However, it is unclear what would cause the most northern tip of the northern extension to have a velocity gradient transverse to that in the extension of 10 \kms.

\section{Summary and Conclusions}\label{concl}

IC 10 is a galaxy with a complex \HI\ morphology and velocity field.  The GBT data reveal an extended disk and a northern extension of \HI.  The VLA data show a southern plume and three spurs.  We have discussed three possible explanations for IC 10's \HI\ morphology and kinematics in this paper:
\begin{enumerate}
\item{The northern extension of \HI\ is partly a bridge between IC 10 and a companion that is at the end of the northern extension.}
\item{IC 10 is an advanced merger where the northern extension is a tidal tail, and the southern plume is a second tidal tail.}
\item{IC 10 is a galaxy still in the process of formation and the northern extension and southern plume are IGM filaments accreting onto IC 10.}
\end{enumerate}

The least likely scenario of the three is the first; the bridge would have likely originated from the potential companion, however there is no evidence of a tidal tail opposite to the bridge associated with the potential companion.  Since tidal interactions are two sided, a tidal tail would have been created on the other side of the companion if there was enough force to create such an extensive bridge, but no such tidal tail is present.  This first scenario cannot explain most of the morphological or kinematic features seen in the \HI\ maps.  The second scenario of an advanced merger suggests that IC 10 has a morphology similar to that of the mice or antennae galaxies.   Since dwarf galaxies are the most abundant type of galaxy at low redshifts \citep{marzke97} and interacting dwarf galaxies, like the SMC and LMC, exist in the Local Group, it is possible that IC 10 has merged with another galaxy. The third scenario of accretion of intergalactic gas filaments would require two separate filaments to be feeding IC 10's disk from opposite directions.  

All of these scenarios suggest that the \HI\ spurs in the VLA data did not originate from IC 10's main disk in the VLA data; implying that IC 10 has likely experienced an external disturbance.  The two most likely scenarios of an advanced merger and gas accretion imply that IC 10 has recently obtained gas which likely has fueled its starburst.  Distinguishing these two scenarios observationally may be possible if there is a faint stellar component associated with the extended tenuous gas seen in the GBT data that extends far beyond the VLA gaseous disk.  In that case the gas would not likely be IGM that is in the process of accreting onto IC 10, but instead gas that has been pulled from the disk of a galaxy.  The lack of a stellar component would not necessarily rule out either scenario since, in the advanced merger scenario, the stellar disk of the companion may have already merged with IC 10's stellar disk and the stellar component of the tidal features may be too faint to be picked up.  

One situation which has not yet been discussed in this paper is the possibility that IC 10 may have interacted with an outside source, such as a galaxy in the M31/M33 group.  Since the \HI\ spurs do not originate from IC 10's main disk, this situation would likely require some combination of the two most likely possibilities: the northern extension may be gas that was gained by IC 10 from the other galaxy and is now accreting onto IC 10, and the southern plume is a tidal tail coming from IC 10's main disk.  However, the closest galaxy to IC 10 within $\pm$150 \kms\ is M31 \citep{hunter04}.  Using the relative distance of 250 kpc and relative velocity of 48 \kms\ \citep[as suggested in][]{hunter04}, the timescale for these two galaxies interacting is \s5 Gyr, yet, the accretion rate time of the southern plume and the northern extension was calculated to be 0.6 and 0.4 Gyr in section~\ref{infalling}.  It therefore does not seem likely that IC 10's northern extension and southern plume could retain their extended structure for 5 Gyr after an interaction with M31.  This result is in agreement with \citet{nidever13}, who suggested that an M31-IC 10 interaction was not likely the cause of the northern extension.  

From the analysis of the GBT and VLA maps of IC 10, it is clear that IC 10 has not been evolving in isolation.  Other BCDs with no apparent companion may also be advanced mergers or galaxies that have accreted IGM.  If that is the case, then these galaxies are examples of how galaxies can obtain more gas for triggering starbursts.  Assuming that these starburst trigger events are rare, a past merger or accretion of IGM support the view that BCDs do not have multiple starbursts over time but rather one epoch of increased star formation in the recent past \citep{schulte01, crone02}.  BCDs in this case may also be useful as analogs of galaxy building in the bottom-up theory.  A larger sample of BCDs that are isolated relative to other galaxies is needed to see if the processes that triggered their starbursts are consistent with a past merger or accretion of IGM, or to see if IC 10 is a special case.

\acknowledgments
We would like to thank Se-Heon Oh for allowing us to use his mass estimate of IC 10.  We would also like to thank the anonymous referee for their helpful comments that improved the presentation of this paper.  Trisha Ashley was supported in part by the Dissertation Year Fellowship at Florida International University.  This project was funded in part by the National Science Foundation under grant numbers AST-0707563 AST-0707426, AST-0707468, and AST 0707835 to Deidre A. Hunter, Bruce G. Elmegreen, Caroline E. Simpson, and Lisa M. Young.  David Nidever was supported by a Dean B. McLaughlin fellowship at the University of Michigan.  This research has made use of the NASA/IPAC Extragalactic Database (NED) which is operated by the Jet Propulsion Laboratory, California Institute of Technology, under contract with the National Aeronautics and Space Administration (NASA).  The National Radio Astronomy Observatory is operated by Associated Universities, Inc., under cooperative agreement with the National Science Foundation.

{}

\end{document}